\input harvmac \noblackbox
\newcount\figno
\figno=0
\def\fig#1#2#3{
\par\begingroup\parindent=0pt\leftskip=1cm\rightskip=1cm\parindent=0pt
\baselineskip=11pt \global\advance\figno by 1 \midinsert
\epsfxsize=#3 \centerline{\epsfbox{#2}} \vskip 12pt
\centerline{{\bf Figure \the\figno:} #1}\par
\endinsert\endgroup\par}
\def\figlabel#1{\xdef#1{\the\figno}}

\def\np#1#2#3{Nucl. Phys. {\bf B#1} (#2) #3}

\def\IR{\relax{\rm I\kern-.18em R}}


\font\cmss=cmss10 \font\cmsss=cmss10 at 7pt
\def\rlx{\relax\leavevmode}
\def\inbar{\vrule height1.5ex width.4pt depth0pt}
\def\IC{\relax\,\hbox{$\inbar\kern-.3em{\rm C}$}}
\def\IN{\relax{\rm I\kern-.18em N}}
\def\IP{\relax{\rm I\kern-.18em P}}
\def\ZZ{\rlx\leavevmode\ifmmode\mathchoice{\hbox{\cmss Z\kern-.4em Z}}
 {\hbox{\cmss Z\kern-.4em Z}}{\lower.9pt\hbox{\cmsss Z\kern-.36em Z}}
 {\lower1.2pt\hbox{\cmsss Z\kern-.36em Z}}\else{\cmss Z\kern-.4em
 Z}\fi}
\def\IZ{\relax\ifmmode\mathchoice
{\hbox{\cmss Z\kern-.4em Z}}{\hbox{\cmss Z\kern-.4em Z}}
{\lower.9pt\hbox{\cmsss Z\kern-.4em Z}} {\lower1.2pt\hbox{\cmsss
Z\kern-.4em Z}}\else{\cmss Z\kern-.4em Z}\fi}

\def\narrowplus{\kern -.04truein + \kern -.03truein}
\def\narrowminus{- \kern -.04truein}
\def\narrowminussub{\kern -.02truein - \kern -.01truein}

\def\frac#1#2{{#1\over #2}}

\def\IZ{\relax\ifmmode\mathchoice
{\hbox{\cmss Z\kern-.4em Z}}{\hbox{\cmss Z\kern-.4em Z}}
{\lower.9pt\hbox{\cmsss Z\kern-.4em Z}} {\lower1.2pt\hbox{\cmsss
Z\kern-.4em Z}}\else{\cmss Z\kern-.4em Z}\fi}
\def\IB{\relax{\rm I\kern-.18em B}}
\def\IC{{\relax\hbox{$\inbar\kern-.3em{\rm C}$}}}
\def\ID{\relax{\rm I\kern-.18em D}}
\def\IE{\relax{\rm I\kern-.18em E}}
\def\IF{\relax{\rm I\kern-.18em F}}
\def\IG{\relax\hbox{$\inbar\kern-.3em{\rm G}$}}
\def\IGa{\relax\hbox{${\rm I}\kern-.18em\Gamma$}}
\def\IH{\relax{\rm I\kern-.18em H}}
\def\II{\relax{\rm I\kern-.18em I}}
\def\IK{\relax{\rm I\kern-.18em K}}
\def\IP{\relax{\rm I\kern-.18em P}}

\font\cmss=cmss10 \font\cmsss=cmss10 at 7pt
\def\IR{\relax{\rm I\kern-.18em R}}

\def\1{{\bf 1}}
\def\3{{\bf 3}}
\def\7{{\bf 7}}
\def\2{{\bf 2}}
\def\8{{\bf 8}}

\def\bbar{{\bar b}}

\def\bbox{{\nabla^2}}

\def\quabla{{\sqcap}\!\!\!\!{\sqcup}}
\def\tria{$\triangleright $}
%

%
%
\def\eqnn#1{\xdef #1{(\secsym\the\meqno)}\writedef{#1\leftbracket#1}%
\global\advance\meqno by1\wrlabeL#1}
\def\eqna#1{\xdef #1##1{\hbox{$(\secsym\the\meqno##1)$}}
\writedef{#1\numbersign1\leftbracket#1{\numbersign1}}%
\global\advance\meqno by1\wrlabeL{#1$\{\}$}}
\def\eqn#1#2{\xdef #1{(\secsym\the\meqno)}\writedef{#1\leftbracket#1}%
\global\advance\meqno by1$$#2\eqno#1\eqlabeL#1$$}


\lref\DuffWD{
M.~J.~Duff, J.~T.~Liu and R.~Minasian, {\it
``Eleven-Dimensional Origin of String/String Duality: A One-Loop Test''},
Nucl.\ Phys. {\bf B452} (1995) 261, hep-th/9506126. }

\lref\rBB{
K.~Becker and M.~Becker, {\it ``${\cal M}$-Theory on Eight-Manifolds,''},
Nucl.\ Phys.\ {\bf B477} (1996) 155, hep-th/9605053.}

\lref\DasguptaSS{
K.~Dasgupta, G.~Rajesh and S.~Sethi,
{\it ``M theory, Orientifolds and $G$-flux''},
JHEP {\bf 9908} (1999) 023, hep-th/9908088. }

\lref\rBHO{E. Bergshoeff, C. Hull and T. Ortin,
{\it ``Duality in the Type II Superstring Effective Action''},
\np{451} {1995}{547}, hep-th/9504081. }
 \lref\rsenorien{A. Sen, {\it ``F-theory and Orientifolds''},
 Nucl. Phys. {\bf B475} (1996) 562,
hep-th/9605150.}

\lref\rstrom{A. ~Strominger, {\it ``Superstrings With Torsion''},
Nucl.\ Phys.\ {\bf B274} (1986) 253.}

\lref\HULL{C.~M.~Hull,
{\it ``Superstring Compactifications With 
Torsion And Space-Time Supersymmetry,''}
In Turin 1985, Proceedings, Superunification and 
Extra Dimensions, 347-375, 29p;
{\it ``Sigma Model Beta Functions And String Compactifications,''}
Nucl.\ Phys.\ B {\bf 267}, 266 (1986);
{\it ``Compactifications Of The Heterotic Superstring,''}
Phys.\ Lett.\ B {\bf 178}, 357 (1986);
{\it ``Lectures On Nonlinear Sigma Models And Strings,''}
Lectures given at Super Field Theories Workshop, Vancouver, 
Canada, Jul 25 - Aug 6, 1986.}

\lref\rDJM{K. Dasgupta, D. P. Jatkar and S. Mukhi, {\it
``Gravitational
Couplings and $Z_2$ Orientifolds''},
Nucl. Phys. {\bf B523} (1998)
465, hep-th/9707224;
J.~F.~Morales, C.~A.~Scrucca and M.~Serone,
{\it ``Anomalous Couplings for D-branes and O-planes,''}
Nucl.\ Phys.\ B {\bf 552}, 291 (1999), hep-th/9812071;
B.~J.~Stefanski,
{\it ``Gravitational Couplings of D-branes and O-planes,''}
Nucl.\ Phys.\ B {\bf 548}, 275 (1999), hep-th/9812088.}

\lref\rBUSH{T. Buscher, {\it ``Quantum Corrections and Extended
Supersymmetry in New Sigma Models''}, Phys. Lett. {\bf B159}
(1985) 127; {\it ``A Symmetry Of The String Background Field
Equations''}, Phys. Lett. {\bf B194} (1987) 59; {\it ``Path
Integral Derivation of Quantum Duality in Nonlinear Sigma
Models''}, Phys. Lett. {\bf B201} (1988) 466.}

\lref\rKKL{E.
Kiritsis, C. Kounnas and D. Lust,
{\it ``A Large Class of New Gravitational and Axionic
Backgrounds for Four-dimensional Superstrings''},
Int. J. Mod. Phys. {\bf A9} (1994) 1361, hep-th/9308124. }

\lref\rgkp{S. Giddings, S. Kachru and J. Polchinski, {\it ``Hierarchies
{}From Fluxes in String Compactifications''}, hep-th/0105097. }

\lref\gates{
S.~J.~Gates,
{\it ``Superspace Formulation Of New Nonlinear Sigma Models,''}
Nucl.\ Phys.\ B {\bf 238}, 349 (1984); S.~J.~Gates, C.~M.~Hull and M.~Rocek,
{\it ``Twisted Multiplets And New Supersymmetric Nonlinear Sigma Models,''}
Nucl.\ Phys.\ B {\bf 248}, 157 (1984); S.~J.~Gates, S.~Gukov and E.~Witten,
{\it ``Two two-dimensional supergravity theories
from Calabi-Yau four-folds,''}
Nucl.\ Phys.\ B {\bf 584}, 109 (2000), hep-th/0005120.}

\lref\rkehagias{A. Kehagias, {\it ``New Type IIB Vacua and
their F-theory Interpretation''}, Phys. Lett. {\bf B435} (1998)
337, hep-th/9805131. }

\lref\GukovYA{
S.~Gukov, C.~Vafa and E.~Witten, {\it ``CFT's from Calabi-Yau Four-folds''},
Nucl.\ Phys.\ {\bf B584} (2000) 69, hep-th/9906070. }
\lref\BeckerPM{
K.~Becker and M.~Becker, {\it ``Supersymmetry Breaking, ${\cal M}$-theory and Fluxes''},
JHEP {\bf 0107} (2001) 038 (2001), hep-th/0107044. }
\lref\DineRZ{
M.~Dine, R.~Rohm, N.~Seiberg and E.~Witten, {\it
``Gluino Condensation In Superstring Models''},
Phys.\ Lett.\ {\bf B156}, 55 (1985).}
\lref\KachruHE{
S.~Kachru, M.~B.~Schulz and S.~Trivedi, {\it
``Moduli Stabilization from Fluxes in a Simple IIB Orientifold''},
hep-th/0201028.}
\lref\FreyHF{ A.~R.~Frey and J.~Polchinski, {\it
``N = 3 Warped Compactifications''}, Phys.\ Rev.\ {\bf D65}
(2002) 126009, hep-th/0201029.}
\lref\CurioAE{
G.~Curio, A.~Klemm, B.~Kors and D.~Lust, {\it
``Fluxes in Heterotic and Type II String Compactifications''},
Nucl.\ Phys.\ {\bf B620} (2202) 237, hep-th/0106155.}

\lref\Vafawitten{ C.~Vafa and E.~Witten, {\it ``A One Loop Test of
String Duality''}, Nucl.\ Phys.\ {\bf B447} (1995) 261,
hep-th/9505053.}

\lref\SethiVW{ S.~Sethi, C.~Vafa and E.~Witten, {\it ``Constraints on
Low-dimensional String Compactifications''},  Nucl.\ Phys.\
{\bf B480} (1996) 213, hep-th/9606122.}

\lref\HananyK{
A.~Hanany and B.~Kol,
{\it ``On Orientifolds, Discrete Torsion, Branes and M Theory''},
JHEP {\bf 0006} (2000) 013, hep-th/0003025.}

\lref\Ganor{
O.~J.~Ganor,{\it
``Compactification of Tensionless String Theories''}, hep-th/9607092.}

\lref\DMtwo{
K.~Dasgupta and S.~Mukhi,
{\it ``A Note on Low-Dimensional String Compactifications''},
Phys.\ Lett.\ {\bf B398} (1997) 285, hep-th/9612188.}

\lref\ShiuG{
B.~R.~Greene, K.~Schalm and G.~Shiu,
{\it ``Warped Compactifications in M and F Theory''},
Nucl.\ Phys.\ {\bf B584} (2000) 480, hep-th/0004103.}

\lref\SmitD{
B.~de Wit, D.~J.~Smit and N.~D.~Hari Dass,
{\it ``Residual Supersymmetry Of Compactified D = 10 Supergravity''},
Nucl.\ Phys.\ {\bf B283} (1987) 165 (1987).}

\lref\PapaDI{ S.~Ivanov and G.~Papadopoulos, {\it ``A No-Go
Theorem for String Warped Compactifications''},
Phys.\ Lett.{\bf B497} (2001) 309, hep-th/0008232.}

\lref\DineSB{M.~Dine and N.~Seiberg,
{\it ``Couplings And Scales In Superstring Models''},
Phys.\ Rev.\ Lett.\  {\bf 55}, 366 (1985).}

\lref\olgi{O. DeWolfe and S. B. Giddings,
{\it ``Scales and Hierarchies in Warped Compactifications
and Brane Worlds''}, hep-th/0208123.}

\lref\hellermanJ{S.~Hellerman, J.~McGreevy and B.~Williams,
{\it ``Geometric Constructions of Non-Geometric String Theories''},
hep-th/0208174.}

\lref\WIP{K. Becker, M. Becker, K. Dasgupta, {\it Work in Progress}.}

\lref\tatar{K.~Dasgupta, K.~h.~Oh, J.~Park and R.~Tatar,
{\it ``Geometric transition versus cascading solution,''}
JHEP {\bf 0201}, 031 (2002), hep-th/0110050.}

\lref\kallosh{K.~Dasgupta, C.~Herdeiro, S.~Hirano and R.~Kallosh,
{\it ``D3/D7 inflationary model and M-theory,''}
Phys.\ Rev.\ D {\bf 65}, 126002 (2002), hep-th/0203019.}

\lref\carlos{R.~Kallosh,
{\it ``N = 2 supersymmetry and de Sitter space,''} hep-th/0109168;
C.~Herdeiro, S.~Hirano and R.~Kallosh,
{\it ``String theory and hybrid inflation / acceleration,''}
JHEP {\bf 0112}, 027 (2001), hep-th/0110271.}

\lref\PolchinskiRR{
J.~Polchinski,
{\it ``String Theory. Vol. 2: Superstring Theory And Beyond''}.}

\lref\sethi{
S.~Sethi,
{\it ``A Relation Between N = 8 Gauge Theories in Three Dimensions,''}
JHEP {\bf 9811}, 003 (1998), hep-th/9809162.}

\lref\senash{
A.~Sen,
{\it ``Orbifolds of M-Theory and String Theory,''}
Mod.\ Phys.\ Lett.\ A {\bf 11}, 1339 (1996), hep-th/9603113.}

\lref\WittenBS{
E.~Witten, {\it ``Toroidal Compactification without Vector Structure''},
JHEP {\bf 9802} (1998) 006 (1998), hep-th/9712028.}

\lref\MeessenQM{
P.~Meessen and T.~Ortin,
{\it ``An Sl(2,Z) Multiplet of Nine-Dimensional
Type II Supergravity Theories''}, Nucl.\ Phys.\ B {\bf 541}
(1999) 195, hep-th/9806120.}

\lref\SenJS{
A.~Sen, {\it ``Dynamics of Multiple Kaluza-Klein Monopoles
in M and String Theory''}, Adv.\ Theor.\ Math.\ Phys.\
{\bf 1} (1998) 115, hep-th/9707042.
}

\lref\BeckerNN{
K.~Becker, M.~Becker, M.~Haack and J.~Louis, {\it
``Supersymmetry Breaking and $\alpha'$-Corrections to Flux
Induced  Potentials''}, JHEP {\bf 0206} (2002) 060,
hep-th/0204254.}

\Title{\vbox{\hbox{hep-th/0209077} \hbox{SU-ITP-02/34}}} {\vbox{
\hbox{\centerline{Heterotic Strings with Torsion}}} }
\centerline{Katrin Becker\footnote{$^1$}{katrin@physics.utah.edu}}
\vskip 0.2cm \vbox{\hbox{\centerline{\it Department of Physics,
University of Utah}} \hbox{\centerline{\it Salt Lake City, UT
84112-0830}}} \vskip 0.2cm \centerline{and} \vskip 0.2cm
\centerline{Keshav
Dasgupta\footnote{$^2$}{keshav@itp.stanford.edu}} \vskip 0.2cm
\vbox{\hbox{\centerline{\it Department of Physics, Stanford
University}} \hbox{\centerline{\it 382 via Pueblo Mall, Stanford
CA 94305-4060}}}

\vskip 0.5in \centerline{\bf Abstract}

In this paper we describe the heterotic dual of the type IIB
theory compactified to four dimensions on a toroidal orientifold
in the presence of fluxes. The type IIB background is most easily
described in terms of an ${\cal M}$-theory compactification on a
four-fold. The heterotic dual is obtained by performing a series
of $U$-dualities. We argue that these dualities preserve
supersymmetry and that the supergravity description is valid after
performing them. The heterotic string is compactified on a
manifold that is no longer K\"ahler and has torsion. These
manifolds have to satisfy a number of constraints, for example,
existence of an holomorphic three-form, size limits, torsional
equations etc. We give an explicit form of the background and
study the constraints associated to them.

\vskip 0.1in \Date{9/02}

\newsec{Introduction}

The standard model has had great success in describing elementary
particles and their interactions. But since the standard model
contains many free parameters we hope that there exists a theory
beyond the standard model capable of explaining quantities like
the pattern of quark and lepton masses or the size of the gauge
hierarchy. String theory is an
excellent candidate to be such a theory since it does not contain
any free dimensionless parameters at all. Instead it has many
scalar fields, the moduli fields, that describe the different
shapes and
sizes of the internal manifold. The parameters of the standard
model should
then ultimately be fixed in terms of the vacuum
expectation values of these fields.
To leading order in the supergravity approximation string theory
contains many degenerate ground states labelled by the moduli fields
since their expectation values cannot be determined.
This is an unattractive situation since it seriously limits the
predictive power of string theory.

The string and ${\cal M}$-theory effective actions contain
quantum gravity corrections which are of higher order in derivatives.
After taking these into account the tensor fields acquire
non-vanishing expectation values. The first example of this
phenomenon were found in \HULL, \rstrom\  for compactifications
of the heterotic string. In the presence of these fluxes
the moduli fields are no longer arbitrary.
Indeed, in \DasguptaSS\
and \GukovYA\ the constraints on the moduli fields
for supersymmetric type IIB compactifications to four dimensions were
obtained by lifting the ${\cal M}$-theory compactifications on
a Calabi-Yau 4-fold found in \rBB\ to ${\cal F}$-theory.

In the context of ${\cal M}$-theory compactifications
on Calabi-Yau 4-folds the fluxes give rise to a supersymmetric background
if
\eqn\oi{
G \wedge J =0 \qquad {\rm and} \qquad G_{(3,1)}=G_{(4,0)}=0,
}
where $G$ is the 4-form of eleven-dimensional supergravity
and $J$ is the K\"ahler form of the internal manifold.
The first equation in \oi\ determines the K\"ahler moduli and
the second equation determines the complex structure.

The supersymmetric backgrounds are not the only solutions of
the equations of motion. Indeed, any self-dual
flux, i.e. any flux that satisfies $G = \star G$,
will solve the equations of motion. So it is
consistent to turn on a flux proportional to the
holomorphic $(4,0)$-form of the Calabi-Yau or proportional to
$J \wedge J$ \GukovYA, \BeckerPM. These additional
vacua have two intriguing properties: they break
supersymmetry and they have a vanishing cosmological constant.
This is certainly
very interesting from the phenomenological point of view.

If we include fluxes,
besides having vacua
with a vanishing cosmological constant and a broken
supersymmetry, it is possible to determine the ground
state of the theory, as we have seen. Moreover in compactifications
with fluxes a non-vanishing warp factor has to be included in
the metric. This warp factor provides one of the few known
mechanisms for naturally generating a large hierarchies of
physical scales. Recently it has been speculated that
the masses generated by the fluxes could even be at the
phenomenological viable TeV scale \olgi.
All these properties are essential in order
to make contact between string theory and the real world.

In this paper we would like to understand if these
properties can also be found in compactifications
of the heterotic string.

In the context of the heterotic string compactified
to four dimensions it is also possible to include
fluxes. But the situation is a bit different than
in type II compactifications. Indeed, if we compactify
the heterotic string on a Calabi-Yau 3-fold
non-vanishing fluxes will always break supersymmetry.
The difference to type IIB compactifications arises because
there is only one ${\cal H}$-field for the heterotic string rather
than two. It is only possible to turn on a flux proportional
to the $(3,0)$-form \DineRZ.
If we start with a manifold that is non-K\"ahler
and has torsion a flux is needed rather than forbidden
to have an unbroken supersymmetry \HULL, \rstrom, \SmitD.
The space-time
metric will no longer be a direct product but we
have to include a warp factor.

In this paper we discuss the background
for a compactification of the heterotic string
on a manifold with torsion. In section 2 we start by
constructing examples of ${\cal M}$-theory
compactifications with fluxes to three dimensions on
$T^8 /{\cal G}$ where ${\cal G}$ is an orbifold group.
 One of the examples is the ${\cal M}$-theory
lift of a compactification on the orientifold $T^6/\IZ_2$ recently
discussed by Kachru, Schulz and Trivedi (KST) \KachruHE. We
discuss the massless multiples that are present in three
dimensions and find  agreement with the field content of the KST
model in one example. In section 3 we lift the ${\cal M}$-theory
compactification on $T^4 /{\cal G} \times T^4 /{\cal G}'$ to
${\cal F}$-theory or equivalently to a compactification of the
type IIB theory to four dimensions. By performing a series of
$U$-duality transformations we find the background for the
compactification of the type I/heterotic string to four
dimensions. We obtain the metric and fluxes for the type
I/heterotic side. In section 4 we consider a concrete example of
an heterotic dual. The six-dimensional internal manifold we get is
no longer K\"ahler and has torsion since on the type IIB side we
started with non-vanishing NS-NS fluxes. These manifolds were
described by Hull \HULL, Strominger \rstrom\ and by de
Wit-Smit-Hari Dass \SmitD\foot{The torsional backgrounds have also
been addressed in a series of papers by Gates et. al \gates.
However they considered a torsion that is closed. This is not
the case we would like to consider here.}. An explicit compact
example was first given in
\DasguptaSS. In the later part of the paper we show in detail that
the background satisfies the torsional constraints that are
imposed by supersymmetry. We conclude with some comments on
various extensions of our idea.

\newsec{${\cal M}$-Theory Compactified on $T^8/{\cal G}$}

\subsec{Type IIB String Theory on the Orientifold $T^6/\IZ_2$}

Recently motivated by \DasguptaSS\
Kachru-Schulz-Trivedi \KachruHE\ and Frey-Polchinski
\FreyHF\
studied a novel type IIB compactification
on the $T^6/\IZ_2$ orientifold in the presence
of NS-NS and R-R 3-form fluxes.
The orientifold action can be denoted by
$\Omega (-1)^{F_L} {\cal I}_6$ where
${\cal I}_6$ stands for the reflection of all
the compactified dimensions.
In this compactification many of the moduli get fixed.
In the presence of 3-form fluxes the superpotential
\eqn\kstsuper{ W = \int G_3 \wedge \Omega, }
is generated. Here $\Omega$ is the unique holomorphic
$(3,0)$-form on $T^6$, and $G_3$ is given in
terms of the NS-NS 3-form $H$ and the R-R 3-form
$H'$ by $G_3= H' - \varphi H$. We denote the
axion-dilaton combination by $\varphi= \tilde \phi
+ i e^{-\phi}$. Observe that $W$ survives on $T^6/\IZ_2$.

The
supersymmetry preserving background follows from minimizing the
superpotential with respect to
the dilaton and the complex structure $\tau_{ij}$ of the torus
\eqn\supeqn{W = 0,\qquad  \partial_{\phi} W = 0\qquad {\rm and} \qquad
 \partial_{\tau_{ij}}W=0. }
Solving \supeqn\ KST showed explicitly
how many of the moduli in this problem pick up some
definite values.

The compactification of the type IIB theory on $T^6/\IZ_2$
can also be studied directly from the ${\cal M}$-theory point of
view where many of the equations get simplified.
In fact, the type IIB model follows
from a simple orbifold of ${\cal M}$-theory in the
presence of $G$-fluxes.
In the next couple of sections we shall elaborate this.
We will show how the
multiplet counting, anomaly analysis etc. result from the
${\cal M}$-theory point of view.

\subsec{${\cal M}$-Theory Lift}

Let us now consider the ${\cal M}$-theory
lift of the type IIB model in some detail.
The base of the 4-fold is $T^6/\IZ_2$ where, as we have seen above,
$\IZ_2$ is an orientifold action. The 4-fold should be a torus
fibration over the base. So we have two possibilities:

\item{1.}
Over each fixed point of $T^6/\IZ_2$ the fiber degenerates as
$T^2/{\cal I}_2$ and everywhere else the fiber is $T^2$.
This would correspond to
the 4-fold $T^8/{\cal I}_8$ where ${\cal I}_n$ is a pure orbifold action
acting on the coordinates $x_i$ of $T^n$ as $x_i \to
-x_i$ for $i = 1,..,n$.

\item{2.}
We could also have the orientifold action of the base to be
$\IZ_2 = \Omega (-1)^{F_L} {\cal I}_2 \times {\cal I}_4$ so that the 4-fold
becomes $T^8/{\cal G} \equiv T^4/{\cal I}_4 \times T^4/{\cal I}_4$.

\noindent For both
cases the Euler characteristic can be related to the perturbative
spectrum of massless multiplets.
In the absence of any fluxes we expect the Euler characteristic to
be given by
\eqn\euler{\chi = 24 n, }
where $n$ is the number of M2-branes.

{}From the ten-dimensional
point of view
this would correspond to the type IIA
string theory on $T^8/{\cal G}$, where ${\cal G}$ can be any of the above
two actions. Under a single $T$-duality of one of the cycles the
$\cal G$ action goes to ${\cal G}' = (-1)^{F_L} {\cal G}$ and we have
the type IIB
theory on $T^8/{\cal G}'$. Here we have $n$ elementary IIB strings
which are required to cancel the 2-form NS sector tadpole \SethiVW,
\Vafawitten.

The Euler characteristic can now be given in terms of $n_1$
(massless states in the NS-R sector) and $n_2$ (massless states in
the R-R sector) as \Vafawitten
\eqn\eulnow{\chi =  2n_1 - n_2,}
where we have used $\int X_8= -\chi/24$.
Here $X_8$ is a polynomial of fourth order in the
Riemann tensor whose definition can be found for
example in \DuffWD.
The values of
$n_1$ and $n_2$ can be argued from the massless spectra in type IIB.

{}For $T^8/{\cal I}_8$ we have
\eqn\ninii{n_1 = -64 \qquad {\rm and} \qquad  n_2 = 256.}
The
negative sign for $n_1$ can be explained as follows. The state
surviving the GSO projection from the left, as well as the ${\cal I}_8$
projection is $({\bf 8_v})_L \times ({\bf 8_s})_R$.
Here $L,R$ denote the left
and right moving sectors of the type II string while ${\bf 8_{v,s}}$ are the
$SO(8)$ vector and spinor representations.
Since $({\bf 8_s})_R$ has
$(-1)^{F_R} = -1$ these contribute a net factor of $-64$.

{}For
$T^8/{\cal G} \equiv T^4/{\cal I}_4 \times T^4/{\cal I}_4$
we have
\eqn\nniii{n_1 =-160\qquad {\rm and } \qquad n_2 = 256. }
There is a small subtlety that we mention at
this point. For the 4-fold $T^8/{\cal I}_8$ the R-R sector twisted
states all have $(-1)^{F_R} = 1$, however the NS-R twisted sectors
are removed because there are no massless states.

Let us now determine the massless multiplets from compactifying the
type IIA theory (or ${\cal M}$-theory) on the above 4-folds.
Let $h_{ij}$ be the
Hodge numbers of the 4-fold. As we know, the metric $g_{MN}$
contributes $h_{11}+ 2h_{31}$ non-chiral scalars, the tensor
fields $B_{MN}$ and  $C_{MNP}$ contribute $h_{11}$ and $2h_{21}$
non-chiral scalars respectively. The spin-$3/2$ particles
come from the gravitinos and their number is given by the Dirac
index of the 4-fold. These ten-dimensional gravitinos
also contribute to the spin-$1/2$ particles,
and their number is given by the
Rarita-Schwinger index. These indices are given by
\eqn\indi{
ind({D\!\!\!\!/}_{3/2}) = 4N - {1\over 6}{\Big (}p_2 - {p_1^2\over
4}{\Big )} \qquad {\rm and} \qquad ind ({D\!\!\!\!/}_{1/2}) =
{N\over 2},}
where $N$ is used to label the supersymmetry as
$(N/2, N/2)$ and $p_i$ are the Pontryagin classes.
Therefore we have the
following multiplets in two dimensions \DMtwo
\eqn\mult{{\Big
(}g_{\mu\nu}, \phi, {N\over 2}\psi_{\mu},
 {N\over 2} \psi{\Big )}~~ \oplus ~~ {2k \over N}{\Big (}{N\over 2} \phi,
{N\over 2} \psi{\Big ), }}
where the first one is the gravity
multiplet and the second one is the non-chiral matter multiplet.

For the 4-fold $k$ is defined as\
\eqn\nplus{ {k\over 2} = h_{11}+ h_{21}+ h_{31}}
If the 4-fold is a
Calabi-Yau manifold then the above expression for $k$ can be
further simplified to
\eqn\kchi{k=
{\chi\over 3} - 16 + 4h_{21}.}

This is however
not the complete spectrum. From \euler\ and \eulnow\ we see that
the consistent vacuum also contains wandering fundamental
strings. The collective coordinates of these fundamental strings
contribute 8 copies of $(\phi, \psi)$ to the total number of
$(N/2,N/2)$
non-chiral matter multiplets. Therefore a consistent vacuum will
have $m$ scalar multiplets where $m$ is given by
\eqn\mgiven{ m =
{2k \over N}   + {2\over 3} { |2n_1 - n_2| \over N}. }

For $T^8/{\cal I}_8$ we have $N=16$ and the Hodge numbers
are $h_{11}= 16, h_{21} = 0, h_{31} = 16$ and
$k = 64$. Therefore we have
\eqn\iiamult{(g_{\mu\nu}, \phi,
8\psi_{\mu}, 8\psi)~~ \oplus ~~ 24
(8\phi, 8\psi), }
massless multiplets in two dimensions. Observe that if
we ignore the contributions from wandering strings then
we find agreement with the multiplets of the type IIB
model considered in \KachruHE.
Indeed, if we compactify the examples considered in \KachruHE\
on $T^2$, so that we would have the type IIB theory
compactified on $T^6/\IZ_2 \times T^2$, then we would
have a total of 65 scalars. This agrees exactly with
the number of scalars found in \iiamult\ if we ignore the
wandering strings.

For the $T^4/{\cal I}_4 \times
T^4/{\cal I}_4$ case we have $N=8$ and
the Hodge numbers are $h_{11} = 40, h_{21} = 0, h_{31} = 40$ and
$k = 160$. Therefore we have
\eqn\iiamullt{(g_{\mu\nu}, \phi,
4\psi_{\mu}, 4\psi) ~~ \oplus ~~ 88(4\phi, 4\psi), }
massless non-chiral
multiplets in two dimensions.

\subsec{Flux Analysis and M2-Branes}

Let us now consider the situation in the presence of fluxes.
${\cal M}$-theory on a 4-fold ${\cal Y}$ can be related to IIB in two
different ways. We can have ${\cal Y} \times T^2$ and shrink the
two torus to zero size. This way we have type IIB on a 4-fold
${\cal Y}$. On the other hand if ${\cal Y}$ itself is a $T^2$
fibration over a base ${\cal B}$ then we get type IIB on a
six-manifold ${\cal B}$.

{}For the case that we consider the type IIB theory compactified
on a 4-fold ${\cal Y}$ the
two-dimensional vacuum preserves chirality and hence induces a vacuum
momentum given by \Ganor, \DMtwo
\eqn\vacmom{L_0 - {\bar L}_0 = {\chi \over 24}.}
This is computed by using the fact that a free periodic boson has
vacuum energy $-1/ 24$ while a free periodic fermion has
energy $1/24$. In the absence of fluxes this vacuum
momentum is related to the O-plane and D-branes charges
of the type IIB theory compactified on ${\cal B}$.

For ${\cal Y} = T^8/{\cal I}_8$
and ${\cal B} = T^6/\IZ_2$ the 4-form of eleven-dimensional
supergravity $G$ will satisfy
\eqn\mtheqn{d \star G =-
{1\over 2} G \wedge G - 4 \pi^2 X_8 - 4\pi^2
\sum_{i=1}^n\delta^8(x - x_i),
}
when ${\cal M}$-theory is compactified on the 4-fold. The Hodge
${\star}$-operator is defined with respect to the warped metric.
We denote the number of M2-branes by $n$ and are
also using the membrane tension $T_3=1$.

The above equation integrates to give the
anomaly condition. Therefore when the type IIB is compactified on the
six-manifold the $X_8$ term
should be related to the number of O3-planes and $n$ will be
correspond to the number of D3-branes.
The $G \wedge G$ term will give rise to the
$H \wedge H'$ term in type IIB.
Therefore \mtheqn\ will
eventually become
\eqn\iibanomaly{
16= n - \int_{\cal B} H \wedge H' . }

For the case when ${\cal Y} = K3 \times K3$ and
${\cal B} = T^2/\IZ_2 \times K3$ the analysis is a little more
elaborate. Going from ${\cal M}$-theory to type IIB
gives us not only
O-planes but also D-branes. In fact, as discussed earlier in many
papers, we get 4 O7-planes
and 16 D7-branes. These D7-branes when wrapped on the other K3
give rise to anti-D3-branes. Details of this have appeared in
\rsenorien. We would like to concentrate on the origin
of the gauge fluxes on the D7-branes. The ${\cal M}$-theory
origin of these fluxes are the {\it localized} $G$-fluxes near the
singularities of $T^4/{\cal I}_4$. To make this a little more precise,
let us concentrate on the region near one orbifold singularity.
This local region will look like a Taub-NUT space. Therefore
${\cal M}$-theory is compactified on $TN \times K3$ where the various
directions are
\eqn\dirtn{\matrix{TN:~~~~-&-&-&-&-&-&7&8&9&10\cr
K3:~~~~-&-&3&4&5&6&-&-&-&-}}
If we denote the metric of $TN
\times K3$ by $g_{a{\bar b}}$, where $a,\bar b$ label the complex
coordinates, and the metric of the flat space-time by
$\eta_{\mu\nu}$, then the metric for the whole system will be warped as
\eqn\lowest{ ds^2 = e^{-\Phi(y)}\eta_{\mu\nu} dx^\mu dx^\nu +
e^{{1\over 2} \Phi(y)} g_{a\bbar} dy^a dy^\bbar ,  }
where $\Phi(y)$ is the warp factor.

The $G$-flux can be expanded as\foot{This equation has also appeared in 
\tatar\ and \kallosh\ in a different context. In \tatar\ this is used to 
understand the supergravity dual of ${\cal N} = 1$ theories and in 
\kallosh, inflationary scenario. It is shown in \kallosh\ that choosing 
$\omega_i$ to be non-primitive, we could still make $G$ primitive and
this could be used to reach a supersymmetric background which is the minima
of the hybrid inflationary-potential \carlos.} 
\eqn\gfluxch{ {G \over 2\pi}
= dz \wedge \omega_1 + d{\bar z} \wedge \omega_2 + \sum_{i = 1}^4
~ F^i \wedge \Omega^i, }
where $i= 1,\dots ,4$ labels the four set of
singularities on the base $T^2/{\cal I}_2$ of $T^4/{\cal I}_4$, $z$ is the
complex coordinate of the fibre and $\omega$ is a 3-form on
the base ${\cal B}=T^2/\IZ_2 \times K3$. As we know, at every fixed points of
the base the fiber is $T^2/{\cal I}_2$ with four fixed points. Therefore
one set of singularities in our notation will have four fixed points.
$\Omega^i$ are the harmonic two-forms on $T^4/{\cal I}_4$.
The zero-modes of the two-forms satisfy the condition
\eqn\zero{
\nabla_{[m}\Omega^i_{np]} = 0. }

The solution of \zero\ can be divided into self-dual and anti-self-dual
parts. The self-dual part, which is a singlet of the holonomy
group $SU(2)$, is covariantly constant tensor and is not
normalisable. We choose the anti-self-dual
part which is normalisable and hence localized at the
fixed points of $T^4/{\cal I}_4$. These localized forms are used to get
gauge fields $F^i_{a{\bar b}}$ on the D7-brane world-volume.
The $X_8$ term
of ${\cal M}$-theory now contributes to the gravitational couplings
on both D7-branes and O7-planes. The warp factor satisfies
\eqn\warpnow{
\eqalign{{\tilde \star}~ \bbox \ e^{3\Phi/ 2}
=  & H\wedge H' +(4 \pi^2 \alpha')^2   \sum_{i=1}^n\delta^{(6)} (x-x_i) +
\cr & { (\alpha' \pi)^2 \over 4} \sum_{j=1}^4 \left[
{\rm tr} ( R \wedge R) - {\rm tr} (F^j \wedge F^j) \right]
\delta^{(2)}( x-x_j), \cr }}
where $n$
is the number of D3-branes and the sum is over four copies of one O7-plane plus
four D7-branes. The factor of a quarter in front of the curvature terms
can be calculated by carefully taking the gravitational couplings of
D7- and O7-planes into account \rDJM.
The Hodge ${\tilde \star}$ and the Laplacian are
with respect to the unwarped metric.

For ${\cal M}$-theory on $T^8/{\cal I}_8$ one can show that there are {\it no}
normalisable harmonic 2-forms at the orbifold singularities.
This is related to the fact that $T^8/{\cal I}_8$ cannot be blown up to a
smooth Calabi-Yau. In other words the blow up operators are
irrelevant rather than marginal. This is consistent because we
only have O3-planes from the singularities and no D-branes\foot{An alternative
way to see this is as follows. To get gauge fluxes we need $h^{1,1} = 256$ but
from Hodge diamond  $T^8/{\cal I}_8$ has $h^{1,1} =16$. All these are from the
untwisted states and therefore constant.}.

Finally let us understand the ${\cal M}$-theory
lift of the exotic O3-planes discussed
in \KachruHE. We shall be brief here and the reader wishing to understand
more should consult the literature \sethi\ \senash\ \WittenBS\ 
\FreyHF\ \HananyK.
In the above discussion, there are
four 2-planes each with charges $-1/ 16$. Combining them we get a
total charge of $-1/ 4$ which is the expected value of a $O3$-plane
in IIB. To get the exotic $O3$-plane one has to combine two 2-planes and
two $OM2^{+}$ planes. These $OM2^{+}$ planes have charges $-3/ 16$ each.
Combining them we get the exotic $O3$ charge of $1/ 4$.

\newsec{A Compactification of the Heterotic String on a Manifold with Torsion}

In this section we will study a specific example of Heterotic
$SO(32)$ string compactification with torsion. We begin with type
IIB theory compactified on an orientifold and perform a series of
$U$-duality transformations to reach our heterotic model. The type
IIB theory that we start off with is compactified on $T^6/\IZ_2$.
The $\IZ_2$ action is $\Omega (-1)^{F_L} {\cal I}_6$ and
${\cal I}_6 = {\cal I}_{897654}$ reverses the all the  tori directions. In the
following note we shall however
assume that the $\IZ_2$ action acts as $T^2/\IZ_2
\times T^4/{\cal I}_4$ where $\IZ_2 = \Omega (-1)^{F_L} {\cal I}_{89}$ and
${\cal I}_4 = I_{7654}$.

We also have a nontrivial $H$ and $H'$ background in type
IIB. However we
should be careful about the choice of the $H$ and $H'$
fields. It
is known that under the orientifold action $\Omega(-1)^{F_L}$
the two $H$ fields  are
transformed by the $SL(2,\IZ)$ matrix
\eqn\mat{M = \pmatrix{-1 &
0\cr 0 & -1}.}
Therefore only those fields which have one of their
components along the torus directions will survive this
projection. Thus our $H$ and $ H'$ backgrounds
have a leg along the $T^2$ parametrized by the $8,9$ coordinates.

The two $T$-dualities along the two circles of $T^2/\IZ_2$ will
convert $\Omega (-1)^{F_L}{\cal I}_{89}$ to $\Omega$ \rsenorien. In other
words it will take our orientifold theory to
type I theory. In what follows we will
specify the exact backgrounds for the type I theory. In the special
case when we start with type IIB with $H =  H' = 0$, the type
I theory that we get by performing two $T$-dualities
 is compactified on $K3 \times T^2$. In the presence of backgrounds
this is no longer the case. On the type I side we get a
six-dimensional manifold which is a non-K\"ahler complex manifold with
vanishing first Chern class \DasguptaSS. And there is a nontrivial
antisymmetric field background on it which we can specify exactly.

But before we go into this we should first see how the couplings
and volumes of the compactified space change under the series of
$U$-duality transformations.

\subsec{The Coupling Constant Maps}

Let us take the type IIB theory compactified on $T^2/ \IZ_2 \times
T^4/{\cal I}_4$.
We take the volumes of $T^2/\IZ_2$ and $T^4/{\cal I}_4$ to be
$\tilde v$ and $v$ respectively. In total the six-dimensional
internal manifold has a volume $V=v \tilde v$. We also define
$\tilde v = R_1 R_2$ where $R_1, R_2$ are the two radii of
$T^2/\IZ_2$.

Under two $T$-duality transformations, which take $R_i \to
\alpha'/R_i$ ($i= 1,2$), we go from the type IIB theory to the
type I theory. The various couplings constants and volumes can now
be specified in terms of type IIB variables:
\eqn\coupI{ g_I= {\alpha' g_B \over  \tilde v}  ,\quad \tilde
v_I={\alpha'^2\over \tilde v}, \quad v_I= v\quad {\rm and} \quad
g_I^{(4)}= {g_B\over \sqrt{v\tilde v}}.}
Here $g_I$ and $g_B$ denote the ten-dimensional type I and type
IIB couplings respectively. By $g_I^{(4)}$ we denote the
four-dimensional type I coupling.

Under an $S$-duality transformation we get the heterotic $SO(32)$
theory. The various couplings and volumes in this theory can again
be written in terms of type IIB variables:
\eqn\couphet{ g_{het}={\tilde v\over \alpha' g_B} , \quad\tilde
v_{het}= {\alpha'\over g_B}, \quad v_{het}= {v\tilde v^2\over
\alpha'^2 g_B^2} \quad {\rm and} \quad g^{(4)}_{het}=
\sqrt{{g_B\over v\alpha'}}.}
In order to get {\couphet} we have used the $SO(32)$
heterotic-type I relations {\PolchinskiRR}
\eqn\bi{ \eqalign{ G_{MN}^{het} & = g_{I}^{-1} g_{MN}^{I}, \cr
g_{het} & =  g_I^{-1}.\cr }}
Observe that the volume $\tilde v_{het}$ is independent of $\tilde
v$ and only depends on the type IIB coupling.

In compactifications of the type IIB theory to four dimensions
with fluxes the dilaton is generically stabilized {\rgkp}.
Therefore the type IIB coupling is $g_B \approx 1$. Let us
consider the large volume limit of the type IIB theory:
\eqn\IIBlimit{
 g_B \approx 1, \quad \tilde v >>1 \quad {\rm and } \quad v>>1.
}
{}From our mapping to type I we find
\eqn\Ilimit{ g_I^{(4)} \to 0 \quad {\rm and } \quad V_I=\alpha'^2
v/\tilde v.}
Therefore the six-dimensional compact type I volume can be
adjusted by picking some definite ratio between the two volumes in
type IIB keeping their individual values large.

For the $SO(32)$ heterotic case we find:
\eqn\hetlimit{ g^{(4)}_{het}\to 0 \quad {\rm and } \quad V_{het}
>>1.}

{}From the above analysis we see that we can use supergravity
analysis to study the various four-dimensional theories.
There is a subtlety related to the large volume limits considered here.
We shall comment on this in later section.

\subsec{Type IIB Orientifold Background}

To discuss the type IIB orientifold background the starting point
is the effective action.
In the absence of sources it is given
by\foot{Throughout the paper we shall always be in the string-frame, except
in sec. (4.2) where we will go briefly to Einstein-frame. It will hopefully
be clear from the text when we switch frames. Also in our notations
${\tilde g}_{mn}$ will be unwarped metric of six compact directions. Type
IIB coupling constant will be denoted by $g_B$ and we shall keep it throughout
the text even though we take examples where $g_B = 1$.
Moreover the Hodge $\star$ will always signify duality wrt
warped metric, and ${\tilde \star}$ wrt unwarped metric $-$
unless mentioned otherwise.}\PolchinskiRR\
\eqn\ci{
S_{IIB}= S_{NS}+ S_{RR }+S_{CS},
}
with
\eqn\actions{
\eqalign{
& S_{NS}  = {1\over 2\kappa_{10}^2}\int d^{10}x \sqrt{-g}e^{-2 \phi}
\left( R + 4 (\partial \phi ) ^2 - {1 \over 2\cdot 3!} H^2
\right),
\cr
& S_{RR}  = -{1\over 4\kappa_{10}^2}\int d^{10}x \sqrt{-g}
\left( F_1^2 + { 1\over 3!} \tilde H_3^2 +
{1 \over 2\cdot 5!}  \tilde F_5^2 \right) ,
\cr
& S_{CS}  = - {1\over 4\kappa_{10}^2}\int D^+ \wedge H \wedge H '.
\cr
}}
The NS-NS and R-R field strengths $H$ and $H'$
are related to the potentials by
\eqn\caii{
H = dB \qquad {\rm and} \qquad H'=dB'.
}
The fields $\tilde H_3$ and $\tilde F_5$ are defined by
\eqn\cii{
\eqalign{
\tilde H_3 & = H' - \tilde \phi H, \cr
\tilde F_5 & = d D^+ - {1 \over 2} B' \wedge H +{1\over 2} B \wedge H'.\cr
}}
The field strength $F_1$ is
given by the derivative of the axion $F_1 = d \tilde \phi$.

The self-duality condition ${\tilde F}_5 =\star \tilde F_5$ cannot
be derived from any known Lagrangian and must be
imposed by hand. In the absence of sources the equation of motion
for $\tilde F_5$ is
\eqn\ciii{ d\tilde F_5= H \wedge H'.}

We will have D-branes and O-planes in the type IIB bulk.
As discussed in section (2.3) in the present example we will have
4 O7-planes and 16 D7-branes, that induce a negative D3-brane
charge.
These
will introduce additional contributions to the effective action
and equations of motion. The equation of motion for $\tilde F_5$
takes the form \rDJM
\eqn\civ{ d\tilde F_5 = H \wedge H' + \rho, }
where
\eqn\cv{\rho= (4 \pi^2 \alpha')^2   \sum_{i=1}^n\delta^{(6)} (x-x_i)
+{ (\alpha' \pi)^2 \over 4} \sum_{j=1}^4 \left[
{\rm tr} ( R \wedge R)_j - {\rm tr} (F^j \wedge F^j) \right]
\delta^{(2)}( x-x_j)
,
 }

In the presence of $H$ and $H'$ backgrounds and sources
the metric in the string frame will have the following form:
\eqn\cvi{ ds^2 =\Delta(y)^{-1}\eta_{\mu \nu} dx^{\mu}dx^{\nu}+
\Delta(y){\tilde g}_{mn}dy^m dy^n,  }
where $\Delta(y)$ is the warp factor. The warp factor and
4-form potential are related by \rkehagias
\eqn\cvii{ D_{\mu \nu \rho \lambda}={ 1\over g_B}\epsilon_{\mu \nu \rho
\lambda}\Delta(y)^{-2} .}
Therefore the equation of motion for $\tilde F_5$ is the
determining equation for the warp factor.

\subsec{$T$-dual Map and Type I Background}

Let us start with our type IIB orientifold model where the orientifold
action has been described earlier. When we make two $T$-dualities
along the two circles of $T^2/\IZ_2$ the operator
$\Omega (-1)^{F_L} {\cal I}_{89}$ changes to $\Omega$ and we get the
type I theory. We will
assume all the fields and the warp factor to be independent of the
torus directions. The $H$ and  $H'$ fields are non-zero only
on the internal space (i.e on $T^2/\IZ_2 \times K3$). Let us begin with
the situation where the torus $T^2$ is not a square torus, i.e
$g_{89} \ne 0$.\foot{Recall that the directions of
$T^2/\IZ_2 \times K3$ are $x^{9,8,...,4}$ and $x^{0,1,2,3}$ are the usual
coordinates of Minkowski space.}
Now we go to the type I theory by
$T$-dualising\foot{ We will assume $4\pi^2 \alpha' = 1$ henceforth.}
 first along the $x^8$ direction and then along
the $x^9$ direction of the $T^2$.

In the next section we shall consider a particular orbifold
example where the complex structures $\tau_{ij}$ of the tori involved are
$i\delta_{ij}$. For such cases
the metric of $T^2/\IZ_2$ can be written in terms of ``flat''
coordinates $w$ in the following way:
\eqn\metric{ds^2 = C~dw
d\bar w,}
 where $C$ is a constant and $w$ is defined as
\eqn\wdeff{
dw =\prod^4_{i=1}~(z - z^i)^{-{1\over 2}} dz.}
However this does not mean that the metric is flat everywhere. At
the four points $z_i$ (which are the fixed points of $T^2/\IZ_2$)
there is a deficit angle of $\pi$. And of course, there is an overall
warp factor. Using the earlier notation of $\Delta$ as the warp factor our
metric is
\eqn\ansatze{\pmatrix{
\Delta^{-1} ~\eta_{\mu \nu}&0\cr
0& \Delta \tilde g_{mn}},}
where ${\tilde g}_{mn}$ is the
unwarped metric of $T^2/\IZ_2 \times K3$.

With the above considerations we can now use Buscher's $T$-duality
rules \rBUSH, \rBHO, \MeessenQM\  to predict
the exact form of type I metric
and the antisymmetric field background\foot{In what follows we use the
conventions of \MeessenQM. In this convention a p-form field will be
defined as $(dA)_{\mu_1 \mu_2 ...\mu_{p+1}}=
(p+1) \del_{[\mu_1}A_{\mu_2..\mu_{p+1}]}$ which symbolically can be denoted as
$(dA)_{p+1} = (p+1) \del A_p$. The convention of \rBHO\ is
$(dA)_{p+1} = \del A_p$. In differential geometry language a p-form will
be defined with $1/p!$ i.e $H = {1\over p!}H_{\mu_1..\mu_p} dx^{\mu_1}
\wedge ...\wedge dx^{\mu_p}$.}.
Since we are taking our torus to be non-diagonal the $T$-dual fields would
be complicated. Let us denote the $T$-dual metric as $g^I_{mn}$ where
$m,n = 4,...,7$ labels the $K3$ directions and $m,n=8,9$ labels the tori
directions. To simplify our formulae let us define two
$2 \times 2$ matrices as
\eqn\gmetrics{g = \pmatrix{
\tilde g_{88}& \tilde g_{89}\cr
\tilde g_{89}& \tilde g_{99}}\qquad {\rm and } \qquad
g^I =\pmatrix{g^I_{88}&g^I_{89}\cr
g^I_{89}& g^I_{99}}, }
where $\tilde g$ is as usual the unwarped metric. The metric in type I theory
is now related to the metric in IIB by the following simple relations:
\eqn\frst{ {\rm tr}~g^I = {1\over \Delta}{\rm tr}~g^{-1} \qquad
{\rm and} \qquad  {\rm det}~g^I = {1\over \Delta^2}~{\rm det}~g^{-1}.}
To find the $g^I_{mn}$ let us define another $2 \times 2$ matrix:
\eqn\Bmatrices{b \equiv b_{(mn)} =\pmatrix{B_{8m}&B_{8n}\cr
B_{9m}& B_{9n}}.}
On the $K3$ base these $B$-fields will form a vector bundle with
globally defined field strength.
Using \Bmatrices\ one can simplify considerably the other components of the
type I metric to:
\eqn\scond{
g_{mn}^I = \Delta \tilde g_{mn} + {{\rm tr}~(b \sigma_4)
~ {\rm det}~(b\sigma_1 + g \sigma_3) + {\rm tr}~(b \sigma_3)~ {\rm det}~
(b \sigma_2 + g \sigma_1)\over \Delta~ {\rm det}~g}, }
where $\sigma_i$ are the Chan-Paton matrices
\eqn\chanpaton{
\sigma_1 = \pmatrix{1&0\cr 0&0}, \qquad  \sigma_2 = \pmatrix{
0&1\cr 0&0}, \qquad  \sigma_3 = \pmatrix{0&0\cr 0&1}, \qquad
 \sigma_4 = \pmatrix{
0&0\cr 1&0}. }
The metric has also picked up off diagonal components which are
related to the background B field by 
\eqn\onea{
g_{8m}^I = { \det (b\sigma_1 + g \sigma_3)\over \Delta ~
\det g}\qquad {\rm and} \qquad g_{9m}^I = { \det (b\sigma_2 +
g \sigma_1)\over \Delta \det g}. }

Analyzing the above equations we find that  we
no longer expect a simple product metric in type I. The metric
is not only warped but also loses its simple product
form.
For the $B^I$ fields we see that it depends on all the type IIB fields like
the 4-form field $D_{mnpq}$, the axion ${\tilde \phi}$ and
the two $B$-fields:
\eqn\bone{\eqalign{
&B^I_{mn}= D_{89mn}^++ 6 B_{[9 m} B'_{n
8]}-4{{\tilde g}_{89} \over \tilde g_{88}} B_{8[ m} B'_{n ] 8} - 2
\tilde \phi B_{9[m }B_{n]8} ,\cr &
B^I_{8m}= - B'_{9m} +  {\tilde \phi} B_{9m}, \qquad
B^I_{9m}= B'_{8m}-\tilde \phi B_{8m}, \qquad B^I_{89}= - {\tilde
\phi}. }}

The
above is therefore the complete background when we do not assume any
approximations. Observe that the metric and the $B$ fields depend
non-trivially on the complex structure. In the next section we
will however consider a slightly simplified background where
\eqn\simplebg{{\tilde g}_{89} = 0, \qquad {\tilde \phi} = 0 \qquad
{\rm and} \qquad D_{89mn}^+ = 0.}
The last equation is trivially true because the 4-form is not self-dual
in the presence of $H$ and $H'$.
The set of equations \frst, \scond, \onea\
and \bone\ simplify drastically by applying \simplebg:
\eqn\sscond{
g_{mn}^I = \Delta \tilde g_{mn} +
{B_{8m}B_{8n}\over \Delta \tilde g_{88}} +
{B_{9m}B_{9n}\over \Delta \tilde g_{99}},
\qquad g_{88}^I ={1\over \Delta \tilde g_{88}},
\qquad g_{99}^I
={ 1\over \Delta \tilde g_{99}}.}
The off-diagonal components of the metric also take a very simple
form. They depend only on the $B$ fields:
\eqn\offdiag{g_{8m}^I ={B_{8m}\over \Delta \tilde g_{88}},
\qquad g_{9m}^I ={B_{9m} \over \Delta \tilde g_{99}}, \qquad g^I_{89}=0,}
together with the $B^I$ fields reducing to
\eqn\btwo{
B^I_{mn}=  6 B_{[9 m} B'_{n 8]}, \qquad
B^I_{8m}=  -B'_{9m},
\qquad B^I_{9m}= B'_{8m},\qquad B^I_{89}=0.}

The set of backgrounds given in \sscond, \offdiag\ and \btwo\ will be
very useful to study the simple orbifold theory in the next
section. The above form of metric can be recast in a more
suggestive way which would clarify the topology of the six-manifold
on the heterotic side. To see this let us first define a vector $A$ as
\eqn\adefin{A = \pmatrix{ dx^8 \cr dx^9}.}

Using this expression we see that metric of the six-dimensional manifold
$K3 \times T^2/\IZ_2$ on the type IIB side takes the form
\eqn\twobmetric{ ds^2 = \Delta {\tilde g}_{mn} dx^m dx^n + \Delta A^{\top}
g A.}

To write the metric on the type I side
we use the set of equations \frst, \scond\ and
\onea. The metric has a fibration structure as
has been discussed earlier \DasguptaSS.
The novelty of the present case is that we
are taking a non-trivial complex structure into account.
This makes the details more interesting. The metric becomes
\eqn\metnowofti{
\eqalign{
ds^2 = \Delta {\tilde g}_{mn}dx^m dx^n +
{A^{\top} g A \over \Delta \det g} & +
{\det  (b\sigma_1 + g \sigma_3)\cdot dx\over \Delta \det g}
\left[ dx^8 + {\rm tr} (b \sigma_4)\cdot dx\right]  \cr
& + {\det (b\sigma_2 + g \sigma_1) \cdot
dx \over \Delta \det g}
\left[ dx^9 +{\rm tr} (b \sigma_3) \cdot dx \right]. } }

Observe that the directions $x^8$ and $x^9$ are non-trivially fibred on the
$K3$ base. The twist in the fiber is governed by the $B$ fields. Let us
consider now the simpler case given by \simplebg, which is the case
considered
earlier in \DasguptaSS.
Under this
assumption the metric simplifies to
\eqn\morphs{
ds^2 = \Delta {\tilde g}_{mn}dx^m dx^n +
{\Delta^{-1}\over {\rm tr} (g \sigma_1)}\left[ dx^8 + {\rm tr}
(b \sigma_1)\cdot dx\right]^2
+{\Delta^{-1}\over  {\rm tr} ( g \sigma_3)}\left[    dx^9 +
{\rm tr} (b \sigma_3)\cdot dx\right]^2. }

Finally
the type I dilaton is related to the
type IIB dilaton by
\eqn\dil{
\phi^I = \phi - {\rm log}~\Delta - {1\over 2} {\rm log} (\det g) .}

 Since $T$-duality is a perturbative symmetry it produces a
consistent background of type I. We will perform various checks in
the later sections to see that this is indeed the case. The type I
background with torsion, as we shall see, has to satisfy many
consistency conditions. In the next section we will take a simple
orbifold model and predict  the various fields.
Also since
$T$-duality preserves supersymmetry,
this will be a supersymmetric background in
four dimensions. As shown in section (3.1) the supergravity
description is valid for both theories, type IIB and
type I. Therefore we expect our heterotic/type I background
to satisfy the conditions found in \HULL, \rstrom\ and \SmitD.

Under an $S$-duality transformation we go to the heterotic
background. In our notation the heterotic string will have a coupling
\eqn\hetcoup{ g_{het} = e^{\phi_{het}} =
e^{-\phi}\Delta~ {\rm Pf} (g), }
where by ${\rm Pf}$ we denote the Pfaffian of $g$.

\newsec{Torsional Background: Detailed Analysis}

In the earlier sections we described
a framework to study a torsional background
which can be {\it derived} from a consistent
background of ${\cal M}$-theory. As we
saw, a warped metric in ${\cal M}$-theory gives us, under
a series of $U$-duality
transformations, a type I background with fluxes turned on.
By performing an $S$-duality
transformation we get a compactification of the
heterotic string on a manifold with torsion.
This background, as briefly alluded to in the introduction, has to
satisfy a number of constraints.
In this section we show that the type I or heterotic
background originating from the ${\cal M}$-theory
compactification on $T^8/{\cal G}$
satisfies all the torsional equations.

\subsec{$T^8/{\cal G}$ Orbifold Example in Detail}

We start with ${\cal M}$/${\cal F}$-theory on $T^8/{\cal G}$,
where ${\cal G}$ is the orbifolding
group. This corresponds to the type IIB theory compactified on the
orientifold $T^2/ \IZ_2 \times T^4/{\cal I}_4$,
 where the $\IZ_2$ action has been defined earlier.

We choose the $G$-flux of ${\cal M}$-theory as
\eqn\Gflux{\eqalign{{G\over 2\pi} =~~  & A
d\bar z^1\wedge dz^2 \wedge d\bar z^3\wedge dz^4 ~+~B dz^1\wedge
d\bar z^2 \wedge dz^3\wedge d\bar z^4 ~+ ~\cr &C d\bar z^1\wedge dz^2
\wedge dz^3\wedge d\bar z^4 ~+~D dz^1\wedge d\bar z^2 \wedge d\bar
z^3\wedge dz^4 + \sum_{i = 1}^4~F^i \wedge \Omega^i,}}
where $z^i$ are the complex coordinates of
$T^8$ and $\Omega^i_{mn}$ are the harmonic 2-forms on
$T^4/{\cal I}_4$. From the form of $G$ we see that it has a constant
part, which is spread over the full 4-fold, and a localized part which is
concentrated at the singular regions. When we go to type IIB by shrinking the
${\cal M}$-theory two-torus (parametrized by $z^4$) to zero size,
the constant part gives rise to
the 3-form fields and the localized part appears as
gauge fluxes on the 7-branes. This form of the
$G$ flux is {\it necessary}
to get a consistent solution in the heterotic theory. As we shall
discuss later, in the absence of the localized flux,
there is probably
no warped
solution. In the next couple of sections we shall however suppress this
contribution and will assume that the $G$-fluxes are constant for our case.
This is not so unreasonable. The contribution from the localized fluxes
come as ${\cal O}(\alpha')$ compared to the other terms in heterotic theory.

Therefore let us reduce the result \Gflux\ along the
$z^4$ direction to go to the type
IIB theory. The constants $A, B, C, D$ are related by the identity
\eqn\ident{ \int_{{\rm vol}} AB + CD = {\chi\over 24}, }
where $\chi$ is the Euler
characteristic of the 4-fold $T^8/{\cal G}$.

Demanding that $G$ is real implies
further constraints on $A, B, C$ and $D$. We set
$A = \bar B$ and
$C = \bar D$ and we introduce the notation
\eqn\notation{
A = \alpha_1 + i \alpha_2 \qquad {\rm and} \qquad C = \beta_1 + i\beta_2.  }
The
anomaly condition \ident\ in the presence of $n$ M2-branes will constrain
$\alpha_i,\beta_j$ by the following relation:
\eqn\constano{ 4(\alpha_1^2 + \alpha_2^2 + \beta_1^2 + \beta_2^2) + n = 24.}
To derive \constano\ we have assumed
\eqn\orient{\int_{\Gamma_x^k} dz^i = \delta^i_k \qquad {\rm and}\qquad
\int_{\Gamma_y^k} dz^i = i \delta^i_k, }
as our periods for a  smooth manifold with $\Gamma_{x,y}^i$ being the
$x, y$ one-cycles. Furthermore, we use the following quantization conditions:
\eqn\quant{\alpha_1 \pm \beta_1 \in 2 \IZ \qquad {\rm and}\qquad
\alpha_2 \pm \beta_2 \in 2 \IZ.}

These quantization conditions take into account
that $G/2\pi$ takes half-integer values when integrated
over all 4-cycles. This also holds for the twisted sectors.
This is a subtle issue and has to be dealt with case by
case for any given orbifold. More details of this have appeared in
\DasguptaSS. Moreover, the quantization condition is modified
from the usual cases
because the orbifold action reduces the volume of the cycles. For
the above case the volume of the 4-cycles is reduced by a quarter.

Let us consider $dz^4 = dx^{10} + \varphi dx^{11}  $, where
$\varphi$ is the axion-dilaton combination which we
have defined in section 3.2.
This decomposition can be used to get the 3-forms on the
base $K3 \times T^2/\IZ_2$. We set
\eqn\usethis{{G\over 2\pi} = H \wedge dx^{10} -  H' \wedge dx^{11}. }
Therefore after
lifting to ${\cal F}$-theory (or type IIB) the $H$ and $H'$ fields are
\eqn\Hfield{\eqalign{
H =& A d\bar z^1\wedge dz^2\wedge d\bar z^3
+ D dz^1\wedge d\bar z^2\wedge d\bar z^3 + \cr
& B dz^1\wedge d\bar z^2\wedge dz^3+
C d\bar z^1\wedge dz^2\wedge dz^3,  \cr
H' =& \varphi (- A d\bar z^1\wedge dz^2\wedge d\bar z^3
- D dz^1\wedge d\bar z^2\wedge d\bar z^3)-\cr
& \bar \varphi( B dz^1\wedge d\bar z^2\wedge dz^3
+ C d\bar z^1\wedge dz^2\wedge dz^3). \cr}}

We observe that all the field components have one of their
legs along the $z^3$ direction.
Here $z^3$ is the complex coordinate of $T^2/\IZ_2$.
To make connection with
the real coordinates described earlier,
we use $z^3 = x^8 + i x^9$. In terms of
complex coordinates the type IIB
metric is given by
\eqn\metriciib{ds^2 = 2( g_{1{\bar 1}} dz^1 d{\bar z}^1 +
g_{2{\bar 2}} dz^2 d{\bar z}^2 +
g_{3 {\bar 3}} dz^3 d{\bar z}^3) . }

Observe that we have used a diagonal complex structure of the tori with
 $\tau_{ij}
= i \delta_{ij}$ for $i,j = 1,2,3$. To see whether this is the case we have
to study the background a little more elaborately. Let us go back to real
coordinates and define more generally
\eqn\deef{ dz^i = dx^i + \tau^{ij} dy^j \qquad {\rm with}
\qquad \tau_{ij} =
{\rm diag}~(\tau_1, \tau_2, \tau_3). }
The background can be determined in terms of a superpotential whose form
is given by \kstsuper. For our case however
it is\foot{Observe that this choice of 
superpotential differs from the usual choice \kstsuper\
by relative minus signs. This 
is because we preferred to choose $-H'$ as our RR three-from. This also
implies that $G_3 = H' + \varphi H$ is a (2,1) form, as it should.}
\eqn\superman{W = (a_0 + \varphi c_0)~ {\rm det}~ \tau -
{\rm det}~\tau~{\rm tr}[(a + \varphi c)^{\top}{\tau}^{-1}] -
{\rm tr}[(b + \varphi d)^{\top} \tau] - (b_0 + \varphi d_0).}

We will consider general diagonal
matrices of $a_{ij}, b_{ij}, c_{ij}, d_{ij}$ and generic $c$-numbers
$a_0, b_0, c_0, d_0$. This is related to an
example considered in section 4.3 of \KachruHE.
To make contact
with the notation of \KachruHE\ we relabel our
coordinates in the
following way
\eqn\coorlabel{(x_4,x_5,x_6, x_7, x_8, x_9) \rightarrow
(x_1, y_1, x_2, y_2, x_3, y_3).}

Therefore our background will be labelled by
\eqn\bgflux{\matrix{
&a_{ij} = {\rm diag}(a_1, a_2, a_3), \hfill
& b_{ij} = {\rm diag}(b_1, b_2, b_3)\hfill \cr
\noalign{\vskip -0.20 cm}  \cr
& c_{ij} = {\rm diag}(c_1, c_2, c_3),\hfill
& d_{ij}= {\rm diag}(d_1, d_2, d_3).\hfill}}
These matrices are used to write the
background NS-NS and R-R 3-forms. We
shall ignore the $\alpha'$ dependence of these forms for the time
being. In the units of $\alpha'$ they are of order one and the
localized
fluxes in \Gflux\ are of order $\alpha'$. The 3-forms are
\eqn\thforare{\eqalign{&H' = a_0 \alpha_0 +
{\rm tr} (a^{\top}\alpha) +{\rm tr} (b^{\top}\beta)
+ b_0 \beta_0, \cr
&H = c_0 \alpha_0 + {\rm tr} (c^{\top} \alpha) +
{\rm tr} ( d^{\top} \beta) + d_0 \beta_0, }}
where $\alpha, \beta$  are defined in eq (2.17) of \KachruHE.
The superpotential
for our case can be written in terms of the above matrices as:
\eqn\superpoten{\eqalign{
W = & (a_0 + \varphi c_0)\tau_1 \tau_2 \tau_3 -
(a_1 + \varphi c_1) \tau_2 \tau_3 - (a_2 + \varphi c_2) \tau_1 \tau_3
 - (a_3 + \varphi c_3) \tau_1 \tau_2 \cr
&- (b_1 + \varphi d_1) \tau_1 -
(b_2 + \varphi d_2) \tau_2
 - (b_3 + \varphi d_3) \tau_3 - (b_0 + \varphi d_0).}}
Let us now assume our background to satisfy
\eqn\satisfy{\
\tau_{ij} = {\rm diag}~( i, i, i)\qquad {\rm and}\qquad
\varphi =  i.}

Minimizing \superpoten\ with respect
to $\varphi$, $\tau_{ij}$ we have the following
set of ten relations between 16 variables:
\eqn\frel{\matrix{
& a_1+a_2+a_3 = b_0, \hfill  & b_1+b_2+b_3 = -a_0,  \hfill \cr
\noalign{\vskip -0.20 cm}  \cr
& c_1+c_2+c_3=d_0 ,  \hfill   &  c_2+ c_3-b_1 = a_0, \hfill \cr
\noalign{\vskip -0.20 cm}  \cr
& a_2+a_3+d_1=- c_0   , \hfill &  c_1+c_2-b_3= a_0,  \hfill \cr
\noalign{\vskip -0.20 cm}  \cr
& a_1+a_2+d_3=-c_0 , \hfill &     d_1+d_2+d_3= -c_0, \hfill      \cr
\noalign{\vskip -0.20 cm}  \cr
& a_1+a_3+d_2=-c_0 , \hfill &  c_1+c_3-b_2= a_0. \cr}}

In the following we would like justify our choice of
complex structure \satisfy\ by showing that it can be derived
from the superpotential \superpoten.
Indeed, the 3-form \thforare, together with the relations \frel,
reproduces the kind of background
we have used in \Hfield. There is a one-to-one
correspondence between the parameters.
The free parameters are given by
\eqn\corep{\matrix{
&c_0 = A+B+C+D ,   \hfill  & d_0 = i(-A-D+B+C), \hfill   \cr
\noalign{\vskip -0.20 cm}  \cr
& c_1= i(-A+D+B-C), \hfill & d_1=-A+D-B+C.  \hfill  \cr }}
The rest of the parameters are determined in terms of
$c_0, d_0, c_1, d_1$ by
\eqn\rest{
\eqalign{
& c_2= b_1= -b_2 = -c_1 , \cr
& c_3 = a_0 = -b_3 = d_0,  \cr
& d_2 = -a_1 = a_2 = -d_1,  \cr
& d_3 = a_3 = b_0 = -c_0. \cr
}}

Including other terms in \Gflux\ we may also choose to cancel
all the anomalies with branes. In general,
the analysis of $T$-dualities
would also apply  in the presence of branes if we delocalize $z^3$
direction. In the presence of D3-branes we will obtain the
type I theory with
5-branes. These 5-branes will eventually be the heterotic 5-branes.
We would like to consider a background for the heterotic theory
without any small instantons. This would mean that the anomalies in type
IIB or
${\cal M}$-theory have to be cancelled by fluxes.

{}From the analysis done above
we see that the only free parameters are $c_0, d_0, c_1, d_1$.
The rest of the parameters
are determined by some choice of these four variables.
But we are not free to choose {\it any} variables.
They are highly restricted
by \quant\ and \constano.
It turns out that there is only {\it one} choice
by which we can cancel anomalies and satisfy the quantization conditions
\eqn\onechoice{
\eqalign{
& c_0=4 ,\qquad d_0= 0, \cr
& c_1=4 ,\qquad d_1=-4.\cr
}}

This implies $A = 2+ i$ and  $C = i$. For the present case
we do not require any wandering D3-branes to cancel the anomaly.
However the above calculations
did not take the other terms of \Gflux\ into account.
The term with
localized fluxes can take fractional coefficients,
so that different values
of $A,\dots,D$ could be generated in our model\foot{In fact this could actually
be used to fix the moduli in \kallosh. There we needed a $D3-D7$ together
and all the others to be far away. The axion behavior could now be chosen in
such a way that we have non-zero value at every point. This non-zero
value should be tuned in a way that most of the $D7$ and the $(p,q)$
$7$-branes are far away. A similar consideration apply to \tatar. We choose
an axion configuration which would tell us that {\it all} the $7$ branes are
far away.}.

For the ${\cal M}$-theory lift of the model discussed in \KachruHE,
i.e for ${\cal M}$-theory compactified on the 4-fold
$T^8/{\cal I}_8$, the
quantization conditions  and anomaly constraint are \DasguptaSS
\eqn\kstquant{
\matrix{
&\alpha_1 \pm \beta_1 \in \IZ \qquad {\rm and} \qquad \alpha_2 \pm
\beta_2 \in \IZ \cr
\noalign{\vskip -0.20 cm}  \cr
& 8(\alpha_1^2 + \alpha_2^2 + \beta_1^2 + \beta_2^2) + n= 16.}}
For $A = 1 + i, C  = 0$ or $A = 1, C = 1$ we can
have no wandering D3-branes in this model.

As discussed in the previous sections we also have
four D7-branes and  one O7-plane
at every fixed point of the base $T^2/{\cal I}_2$.
The metric
$g_{3 {\bar 3}}$ should not be flat as we expect backreactions from
the branes and planes.
In the general case, if we move one
of the 7-branes in our system slightly then
all the O-planes become $(p,q)$
7-branes under non-perturbative corrections. We shall not repeat the
details of this mechanism here as this has been discussed extensively
by Sen \rsenorien. For our case \satisfy\ tells us that we have a charge
cancelled situation ${\tilde \phi} = 0$ and therefore those moduli are
fixed. Thus the metric will be flat everywhere except at the
singular orbifold points.
{}From the ${\cal M}$-theory point
of view the $X_8$ term is responsible for the $D7$- and
O7-planes. For this case
\eqn\xeight{X_8 = {3\over 32} \sum_{z^i, w^j}~\delta^4(z - z^i)
\delta^4(w - w^j),  }
where $z^i, w^j$ are the fixed points of $T^4/{\cal I}_4
\times T^4/{\cal I}_4$.

{}From equation \Hfield\
it is easy to determine the $B$ and
$B'$ backgrounds. They are:
\eqn\Bfield{\eqalign{
B =& A\bar z^1 dz^2\wedge d\bar z^3
 + B z^1 d\bar z^2\wedge dz^3 - C z^2 d\bar
z^1\wedge dz^3  - D \bar z^2 dz^1\wedge d\bar z^3, \cr
B' =&
 {i\over g_B}\left(- A\bar z^1 dz^2\wedge d\bar z^3
+ B z^1 d\bar z^2\wedge
dz^3 -  C z^2 d\bar z^1\wedge dz^3 + D \bar z^2
dz^1\wedge d\bar z^3\right).   \cr}}
Observe that these $B$ fields are not globally defined. To describe these
fields we have to use patches on the six-manifold. $H$ however is globally
defined.

$T$-dualising to type I, the non zero fields are the metric $g^I_{a{\bar b}}$
and $B^I$ fields. The metric takes the form
\eqn\formofmet{ ds^2~ = 2g^I_{a{\bar b}}~dz^a  d{\bar z}^b. }
The type I metric can be expressed in terms of the type IIB metric and the
NS-NS $B$ field. Under $T$-duality the $B$ field dissolves in
the metric if it has components along the $T$-dual directions.
{}From \Bfield\ we see that this is the case. Therefore
\formofmet\ takes the form
\eqn\sixmetric{\eqalign{
ds^2 & = 2 g_{1{\bar 1}} dz^1 d{\bar z}^1 +
2 g_{2{\bar 2}} dz^2 d{\bar z}^2 \cr
&+ {1\over  2g_{3 {\bar 3}}}
\Big(  dz^3 + 2 {D {\bar z}^2} dz^1 -
 {2A {\bar z}^1} dz^2 \Big )
\Big( d{\bar z}^3 + {2C z^2} d {\bar z}^1  -
{2B z^1} d {\bar z}^2\Big ). }}
As can be easily seen all other components of the metric are zero for the
case that we are considering. The above metric \sixmetric\ gets
further simplified by imposing the condition that
the $G$-fluxes should be real. It takes
the following form
\eqn\metform{ds^2 =  2 g_{1{\bar 1}} dz^1 d{\bar z}^1 +2 g_{2{\bar 2}}
dz^2 d{\bar z}^2 + {1\over 2 g_{3 {\bar 3}}}{\big| }dz^3 +
{2D {\bar z}^2} dz^1 -{2A {\bar z}^1} dz^2 {\big|  }^2.}
The $B^I$ field of type I can be easily found from \btwo\ using the set of
fields \Bfield\ forming a vector bundle. It is given by
\eqn\BIone{
B^I={ 1\over g_B}\left( A \bar z^1~dz^2 \wedge d{\bar z}^3+
B z^1~d{\bar z}^2 \wedge dz^3
-C  z^2~d{\bar z}^1 \wedge dz^3
-D \bar z^2~dz^1 \wedge d{\bar z}^3\right) . }
Indeed, from \Bfield\ one can easily see that $B_{m8}=-g_B B^{'}_{m9}$
and $B_{m9}=g_B B'_{m8}$.
This implies that the only non-vanishing components
have one leg in the $z_3$ or $\bar z_3$ direction.
Finally the type I coupling is related to the type IIB
coupling via
\eqn\couptypeI{g_I = {g_B\over 2 g_{3\bar 3}}.}

{}From the metric components we see that type I theory is
compactified on a non trivial six-dimensional manifold.
The geometry of the
space is clear from \metform.
The $z_3$ and ${\bar z}_3$ directions are fibred in a specific way over the
base. The twist of the fiber torus originates entirely from
NS-NS B field of the type IIB theory.
This is a
solution with torsion and therefore there are lots of constraints
on it. In the next section we will discuss these
constraints and show that our solution does indeed satisfy all the
constraints.

\subsec{Analysis of the Torsional Constraints}

In the Einstein frame a type I/heterotic background is
supersymmetric when it satisfies the conditions\foot{In this section
$\phi \equiv \phi_{het}$ unless mentioned otherwise.}
\eqn\susy{\eqalign{&
\delta \psi_M = \nabla_M \epsilon +
{1\over 48} e^{-{\phi/ 2}}
(\Gamma_M {\cal H} - 12 {\cal H}_M) \epsilon =0 , \cr
&\delta \lambda =  \Gamma^N \nabla_N \phi \epsilon -
{1\over 6}e^{-{\phi/ 2}} {\cal H} \epsilon=0, \cr
&\delta \chi = e^{-{\phi/ 4}} F_{MN} \Gamma^{MN} \epsilon = 0,  }}
for some Majorana-Weyl spinor $\epsilon$. With indices $M,N$ we label
the ten-dimensional coordinates.
Here $\chi$ is the gluino, $\psi_K$ is the gravitino, $\lambda$ is the
dilatino and $\cal H$ is the 3-form field satisfying\foot{
We differ here, for
example, from \rstrom\ by a factor of 2. This factor is
important to compare type IIB and type I theories. It is also important to
determine the gravitational couplings on O-planes.
For details see \rDJM.}
\eqn\heq{
d {\cal H} =  {1\over 2}\left[p_1(R) - p_1(F)\right] =
  {1\over 16 \pi^2}( {\rm tr} R \wedge R -{1 \over 30}
{\rm Tr} F \wedge F).}
We have also defined ${\cal H} \equiv {\cal H}_{MNP}\Gamma^{MNP}$
and ${\cal H}_M \equiv {\cal H}_{MNP}
\Gamma^{NP}$.

The choice of
coordinate system used in \susy\ is not appropriate
because
the metric is not the one that
is used to measure distances. There
is a more convenient coordinate system that is
obtained by transforming the fields in the
following way
\eqn\rescale{
g_{MN} \to e^{\phi / 2} g_{MN},\qquad
 \epsilon \to e^{\phi / 8}
\epsilon,\qquad  \lambda \to e^{-{\phi /8}}\lambda,\qquad  \psi_M \to
e^{\phi / 8}(\psi_M - {1\over 2} \Gamma_M \lambda)}
We denote the resulting heterotic metric by $ G_{m\bar n}$. It is
related to the previous type I metric by
\eqn\hetmetric{
G_{m\bar n}=   {2 g_{3 \bar 3} \over g_B}~ g_{m \bar n}^I.
}
Here we have used that the heterotic coupling is related to
the type IIB coupling by
\eqn\coupling{
g_{het} = e^{\phi}= {2 g_{3 \bar 3} \over g_B}.}

For the particular case of compactification with a maximally
symmetric space-time, unbroken supersymmetry implies that the
warp factor is proportional to the coupling.
It also requires the space-time to have vanishing
cosmological constant \rstrom. To
see this observe that for the choice of metric given in
\rescale\ the covariant derivative is shifted by
\eqn\covderiv{\nabla_M - {1\over 8}   \Gamma_{MN}
\partial^N \log \Delta. }
This implies that the external component of the
gravitino supersymmetry transformation law is
given by
\eqn\fourdim{ \delta\psi_{\mu} = {\nabla}_{\mu} \epsilon - {1\over 8}
{\Gamma_{\mu \nu}}  \partial^\nu ({\rm log} \Delta - \phi)= 0. }
The above conclusions follow by contracting \fourdim\ with
$\Gamma^{\mu\rho} \nabla_{\rho}$.

Our six-dimensional compact manifold is complex with fundamental form
\eqn\cplx{J =i  G_{a {\bar b}} ~dz^a \wedge d {\bar z}^b.}
If $dJ=0$ the manifold is K\"ahler and $J$ is the K\"ahler form.
The complex structure is constructed from
the positive chirality spinor on the six-manifold,
and is $\cal H$-covariantly constant
\eqn\hcovar{
2 { \nabla}_m J_n^{~p} -
{\cal H}_{qm}^{~~p}J_n^{~q} - {\cal H}^q_{~mn} J_q^{~p} = 0.}
Multiplying the above equation by the complex structure
gives us the following important relation
between the background 3-form
and the fundamental form \HULL\ \rstrom\ \SmitD:
\eqn\consback{{\cal H} = i(\bar \del - \del) J, }
which can be expressed in terms of components as
\eqn\compon{
{\cal H}_{\bar m \bar n p}= -2 G_{p[\bar m, \bar n]}
\qquad {\rm and} \qquad
{\cal H}_{m n \bar p}= -2 G_{\bar p[m, n]}.}
Observe that the relation can now be written
purely in terms of type I variables.
The type I metric of course involves the
warp factor in a complicated way as we have seen
in earlier sections. The way we
have derived this differs from \rstrom\ by a factor of 2.

By rescaling the fields as in \rescale\
the set of equations in \susy\
becomes dilaton free. This implies
that the killing spinor equation becomes
\eqn\killing{({\nabla \!\!\!\!\slash}~\phi- {1\over 6}{\cal H})\epsilon = 0,}
which together with \compon\ gives rise to an equivalent relation
between dilaton and ${\cal H}$-field
\eqn\anrel{{\cal H}_{ \bar m n {\bar p}}G^{n{\bar p}} = - \nabla_{\bar m}\phi.}
In general the internal six-dimensional manifold will be non-K\"ahler.
The shift from K\"ahlerity is given precisely in terms of the warp factor as
\eqn\nonkahler{d^{\dagger}J = i (\partial - {\bar \partial})~ {\rm log}\Delta.}
Therefore when $\Delta = 1$ or constant dilaton we have our usual Calabi-Yau
compactification.

 In the type IIB background most of the moduli fields of the internal
manifold are fixed.
{}From our earlier
analysis \satisfy\ we see that $g_B = 1$. However,
we will retain the coupling in all equations.
Our final results will be independent of the
choice of the coupling constant.

To verify the constraints we need the components of
the heterotic metric. It is easy to extract from \sixmetric\
and it is given by
\eqn\gone{
G_{a \bar b}= \pmatrix{
G_{1 \bar 1} & G_{ 1 \bar 2} & G_{1 \bar 3} \cr
\noalign{\vskip -0.20 cm}  \cr
G_{2 \bar 1} & G_{ 2 \bar 2} & G_{ 2 \bar 3}\cr
\noalign{\vskip -.20 cm}\cr
G_{3 \bar 1} & G_{ 3 \bar 2} & G_{3 \bar 3} \cr }
=
{ 2\over g_B} \pmatrix{
g_{1\bar 1}g_{3\bar 3}+CD z^2 \bar z^2  & -B Dz^1 \bar z^2 &  D\bar z^2/2 \cr
\noalign{\vskip -0.20 cm}  \cr
-AC \bar z^1 z^2 & g_{2\bar 2}g_{3\bar 3}+A B z^1 \bar z^1 & - A \bar z^1/2 \cr
\noalign{\vskip -0.20 cm}  \cr
 C z^2/2 & - B z^1/2 & 1/4 \cr}}
There are a set of $24$ equations which follow from \compon.
In the following we will be demonstrating that these
equations are consistent.

Using the duality relations derived in the previous sections we
find the following equations
\eqn\solone{
{\cal H}_{ 2 3 \bar 1} = {C\over g_B} =
-2 G_{\bar 1 [2,3]}\qquad {\rm and} \qquad
{\cal H}_{ 1 3 \bar 2} = -{B\over g_B} =
-2 G_{\bar 2 [1,3],  }}
which are consistent. Observe that since the results
are written in terms of the field strength,
they are globally defined over the
full six-manifold.
The other non-vanishing components are
\eqn\soltwo{
{\cal H}_{ 1 2 \bar 1} =0 \qquad {\rm and} \qquad
 G_{\bar 1 [1,2]}=
 {1 \over g_B} {\del\over \del z^2}( g_{1\bar
1}~g_{3\bar 3}) +{ CD \over g_B}  \bar z^2.}
While the other component is
\eqn\solthree{
{\cal H}_{ 2 1 \bar 2} =0 \qquad
{\rm and } \qquad  G_{\bar 2 [1,2]}=
-{1 \over g_B} {\del\over \del z^1}( g_{2\bar 2}~g_{3\bar 3}) -
{ A B\over g_B}   \bar z^1, }
and their complex conjugates.

If the identity \consback\ has to
hold then we expect \soltwo\ and \solthree\ to vanish.
Therefore we expect
\eqn\check{
{\partial \over
\partial z^2}(g_{1\bar 1}~g_{3\bar 3}) +CD \bar z^2 =
{\partial \over \partial z^1}(g_{2\bar
2}~g_{3\bar 3}) + A B \bar z^1 =0 .}
The metric of the internal space in the type IIB theory
is basically the flat metric on a
square torus $T^6$ multiplied by a warp factor $\Delta$ as shown
in \ansatze. After taking the warp factor dependence into
account \check\ become
linear equations for the
warp factor that take the form
\eqn\warpdete{
{\partial \Delta^{2}\over \partial z^2} +CD \bar z^2 =
{\partial \Delta^ 2\over \partial z^1}+ A B \bar z^1 =0 .}
These linear equations have an origin in the
type IIB side of the duality. Indeed from the definition
and self-duality of $\tilde F_5$ we obtain the equation
\eqn\lineqiib{
 \star ~
 d D^+ = {1\over 2}( B \wedge H' -B ' \wedge H).}
We have also seen that the condition for unbroken supersymmetry
implies a relation between $D^+$ and the warp factor. This relation
implies that \lineqiib\ takes the form
\eqn\newlin{
\star ~ d  \Delta^{-2} ={g_B\over 2} (
 B \wedge H' - B' \wedge H).
}
The Hodge $\star$-operator in \lineqiib\ and
\newlin\ is defined with respect to the ten-dimensional
metric. In order to compare with \warpdete\ we have
to transform this to the Hodge operator with respect to
the flat metric which we denote by $\tilde \star$.
The result is
\eqn\newlini{
{\tilde \star} ~d \Delta^{2}=i \left(
A B {\bar z}_1 dz^1 \wedge d z^2
\wedge d{\bar z}^2 \wedge dz^3 \wedge d{\bar z}^3 +
CD{\bar z}^2 dz^2 \wedge dz^1 \wedge d{\bar z}^1 \wedge
dz^3 \wedge d{\bar z}^3-c.c. \right). }
After expanding in components we obtain exactly \warpdete.
This explains the type IIB origin of the constraints \compon.
Note that \newlini\ also implies that the warp factor
is $z_3$ and $\bar z_3$ independent. This is
consistent with our assumptions.

By acting with the adjoint of $d$ on \newlini\
it is easy to obtain the quadratic warp factor equation
\eqn\warpeq{\bbox \Delta^2 = -2(A B + CD) = - 2~(\alpha_1^2 +
\alpha_2^2 + \beta_1^2 + \beta_2^2), }
which from the ${\cal M}$-theory point of view takes the form
\eqn\warpM{\bbox ~ e^{3 \Phi/ 2} = - {1\over 2}\star~ (G\wedge G) =
- 2~ (\alpha_1^2 +\alpha_2^2 + \beta_1^2 + \beta_2^2).}
We have used the fact that the volume of the fundamental
domain is reduced by
$1/ 4$.
The equations \warpeq\ and \warpM\ are identical if
\eqn\checko{\Delta^2 = e^{3 \Phi/ 2}}
which is indeed the case because it is required
by supersymmetry \rBB.

Observe that in the two equations
above \warpeq\ and \warpM\ we did not
incorporate the delta function source terms. It is straightforward to
incorporate these in \warpM\ but it is not so easy to see how they
appear in
\warpeq. This problem is partly related to the fact that we have taken
$g_{3 {\bar 3}}$ to be a flat metric. Because
of the presence of orbifold points
the flat choice would actually not really work.
To get the exact form of
$g_{3 {\bar 3}}$ one has to solve Einstein's equation in the presence of
fluxes. However, instead of
doing this we can {\it still} view $g_{a {\bar b}}$ to be flat at all points
and incorporate additional delta function contributions by hand. In other
words, if we call the actual Laplacian, which takes into account the
singularities, $\quabla$, then the relation between the
two Laplacians is
\eqn\actualoper{ {\quabla}~ =~ \nabla^2~ +~
{\rm x}, }
where ${\rm x}$ takes the singularites into account. From dimensional
analysis we immediately see that ${\rm x}$ is of order
$L^{-2}$, where $L$ is the
typical length scale on the six-manifold. The singularities should
contribute delta functions to ${\rm x}$, and thus the most general form for
${\rm x}$ is
\eqn\xform{ {\rm x} = N~\sum_{z^i, w^j}
{\delta^2(z - z^i)~\delta^4(w - w^j) \over \Delta^2 },}
where $N$ is a number that could be determined from actually solving the
gravity equations and taking the gravitational effects of
D7-branes and O7-planes into account.
Combining \xform , \actualoper\ and \warpeq\ we get
\eqn\warpeqii{
{\quabla} ~\Delta^2 = - 2 (\alpha_1^2 + \alpha_2^2 +
\beta_1^2 + \beta_2^2) +
N~\sum_{z^i, w^j}~\delta^2(z - z^i)~\delta^4(w - w^j), }
and using \xeight\ we can confirm \checko.

We should mention the following important point.
In deriving the type I background we have assumed that all
the fields are independent of the $z^3, {\bar z}^3$ directions.
However, \warpeqii\
do not reflect this fact. It is of course true that the correct warp
factor equation should not involve delocalization, but
the $T$-duality rules we have used are
not attuned to this. A more general $T$-duality rule is required to
get the background that includes the delta function sources.
But to the lowest order in $\alpha'$ we have obtained
the correct answer and we have verified that it is consistent.

For the simpler case when we ignore the backreaction from
D7-branes we can express the solution of \warpeqii\
in terms of the Green's functions of the
compact space $K$.
Because of the compactness of the internal manifold
there exists an IR cutoff.
This IR cutoff appears as the volume factor
in the Green's function \ShiuG\
\eqn\greenf{{\quabla}~K (z,z^i,w, w^j) =
\delta^2(z - z^i)~\delta^4(w - w^j) - {1\over {\tilde v} v}, }
where ${\tilde v}$ and $v$ have been defined in section 3.2.
The backreaction from the
D7-branes becomes important when the charges are not cancelled locally.
The localized delta function source that has appeared in \warpeqii\
originate from the $X_8$ term of ${\cal M}$-theory.
As we discussed in \Gflux, there is
yet another localized source which is responsible for the gauge fields on
the D7-branes. We have been neglecting that term but it also contributes
to $H$ and $H'$ in \Hfield. The contribution is such that
\eqn\hweh{{\tilde \star}~(H \wedge H') =
4 (\alpha_1^2 + \alpha_2^2 +
\beta_1^2 + \beta_2^2)
+ \sum_{i = 1}^4{\tilde \star}~{\rm tr}~(F^i \wedge F^i)~ \delta^2(z - z^i).}
We are working at the supergravity level in which we do not
expect to see non-abelian gauge fields. But we
have included a trace in front of the gauge fields in \hweh\
because we will argue later that
non-abelian gauge fields should appear after taking wrapped
M2-branes into account. The non-abelian gauge symmetry we
expect for our model is $D_4^4$. Indeed, we have four
points on the base with one O7-plane and 4 D7-branes at every point.
The D7-branes generate an $U(4)^4$ gauge symmetry which is broken
by the O7-planes to an $SO(8)^4= D_4^4$ symmetry.

For the orbifold case the curvature are localized at the fixed points.
This is apparent from \warpeqii\ and \xeight. Therefore
the equation of warp factor will modify from \warpeqii\ to
\eqn\warpyy{{\quabla}~ \Delta^2 =
-2(\alpha_1^2 + \alpha_2^2 + \beta_1^2 + \beta_2^2)+
\sum_{i,j}\delta^2(z - z^i)
\left[ N~\delta^4 (w - w^j)
- {\tilde \star}~ {\rm tr}~(F^j \wedge F^j)\right].}
Comparing with \greenf\ we seem to be missing
some volume factors. The reason for this is
that we have taken unit volumes throughout.

Let us consider now a
simpler situation where the instanton term is a delta function. The warp
factor can actually be derived for this case following the calculations
of \ShiuG. It is given by
\eqn\warpsolution{ \Delta^2 = c_0 + c_1~ K (z,z^i,w, w^j), }
where $c_0, c_1$ are constants and $K$ is the same Green's function
as in \greenf.
In the limit in which the size of the six-manifold is very large
i.e. when $|z|, |w| \to \infty$, we have
\eqn\largesize{ \Delta = {\sqrt c_0}, \qquad  \Gamma^c_{{\bar b}c} = 0
\qquad {\rm and} \qquad dJ = 0, }
implying that the internal manifold is Ricci-flat
and K\"ahler.
Since we also have a vanishing first
Chern-class, \largesize\ would imply that at
large radii these manifolds are
Calabi-Yau 3-folds. For the case that the radius is a
parameter of these string compactifications, quantum effects will prefer
large radii and hence Calabi-Yau manifolds, as has
been argued by Dine and Seiberg in \DineSB.
This would be contradictory! Non-vanishing {\it total} fluxes like
the ones we have in the present model can not be
supported by Calabi-Yau 3-folds. Indeed,
when the heterotic string
is compactified on Calabi-Yau 3-folds fluxes are not allowed
if we want to preserve some supersymmetry.
Therefore we would find that these compactifications
are inconsistent.
This puzzle could be resolved if the solutions we
constructed would contain a scale $\alpha'$ which could be used to
actually fix the radius. Therefore these manifolds
will not have a large radius limit. Some of these manifolds have also been
studied recently in \hellermanJ.

Up to now we have mostly ignored the localized fluxes appearing in
\Gflux. In the following we would like to study the
condition for unbroken supersymmetry on those fluxes. The generic
supersymmetry condition for $G$-fluxes
is given in \oi. For a choice of K\"ahler form
$J$ it is easy to see that the constant part of \Gflux\
satisfies \oi. For the
localized piece of $G$ we find
\eqn\locpiece{
 G_{a \bar b c \bar d}J^{a \bar b} =
 i ~ \sum_{i = 1}^4 ~ g^{a \bar b} F^{i}_{a \bar b}
(z^1, z^2, {\bar z}^1, {\bar z}^2)
~ \Omega^{i}_{c {\bar d}}(z^3, z^4,
{\bar z}^3, {\bar z}^4) = 0.}
Here we have used that $\Omega \wedge J=0$ because of the
anti-self-duality of $\Omega$.
The form $\Omega$ is a ($1,1$)-form and not a ($2,0$)- or
($0,2$)-form because as discussed in \zero\ it is anti-self-dual.
In fact, for $T^4/{\cal I}_4$ sixteen of
these localized forms contribute to $h^{1,1}$ in addition to the already
existing four ${\cal I}_4$ invariant forms $dz^i \wedge d{\bar z}^j$
to complete
the twenty ($1,1$)-forms.
Also since we are performing $T$-dualities
the fields $F_{a {\bar b}}$ and
$g_{a {\bar b}}$ are independent of the ($z^3, z^4$) directions.
We do not need an explicit representation of
$\Omega$, but to extract information from
\locpiece\ we need the orthogonality condition
\eqn\orthocond{\int_{K3}\Omega_{c {\bar d}}^{i}~
\Omega_{e {\bar f}}^{j} =
(16 \pi)^2 \delta^{ij}~\delta_{ce}~\delta_{{\bar d}{\bar f}},}
where we have set the mass parameter on the
right hand side of \orthocond\ equal to 1 (see for example \SenJS).

Integrating \locpiece\ over the 4-manifold $K3$
and using the relation \orthocond\ we obtain the
condition for unbroken supersymmetry satisfied by the
gauge fluxes
\eqn\susycondonfluxes{ g^{a {\bar b}}F^{i}_{a {\bar b}} =0.}
This is of course the expected primitivity condition for the gauge fields on
the D7-branes. This serves as another confirmation that the
localized fluxes are the 7-brane gauge fields.

The above analysis only gave us the gauge fields that are in the Cartan
subalgebra of
$D^4_4$. The couplings and Lagrangian of the ten-dimensional
fields involved in the
decomposition \Gflux\ and \usethis\ can be worked out easily. By plugging
in the value of $G$-flux \Gflux\ one can show that
the Lagrangian of ${\cal M}$-theory decomposes as
\eqn\decomofMth{
\eqalign{\int G\wedge \star G ~ & \rightarrow ~ \int d^{10}x ~(
g_B^{-2} |H|^2 + |H'|^2 {)} ~+~
\sum_{i = 1}^4~\int d^8 \sigma ~|F^i|^2 \cr
 \int  C \wedge G \wedge G ~ & \to ~ \int  ~ D^+ \wedge H
\wedge H' ~+~ \sum_{i = 1}^4 \int d^8 \sigma ~D^+ \wedge F^i \wedge F^i.}}
The first terms appear in the type IIB bulk
and the second terms are interactions
on the D7-brane world-volume.
The $C \wedge X_8$ term  gives rise to the couplings on the D7-branes
and O7-planes only and it gives no contributions to the
bulk interactions. We have used the orthogonality
condition for the components of $\Omega$
to get the interactions of the D7-brane world-volume
gauge fields. As we see, this analysis only gives the {\it abelian} part
of the gauge group (i.e the Cartan subalgebra).
By taking wrapped M2-branes on the vanishing
cycles at the fixed points into account we can argue for the
complete non-abelian symmetry of our model.
These M2-branes
are wrapping 2-cycles of the 4-fold and
are therefore different from the anomaly cancelling
M2-branes in the $(2+1)$-dimensional space-time.
Using $U$-duality it
should be possible to see a similar kind of
relation for the heterotic gauge
fields. We would need a similar ($1,1$)-form for our
six-manifolds. We postpone the details of this analysis to the future.

Finally, as discussed by Strominger, the six-manifold supports a unique
holomorphic $(3,0)$-form. This form is in the middle dimensional
cohomology.
Let us recall the metric of the six-manifold again (we take $g_B = 1$ for 
simplicity here):
\eqn\metagain{ds^2= 4g_{3 {\bar 3}} g_{1{\bar 1}} dz^1 d{\bar z}^1 +
4 g_{3 {\bar 3}} g_{2{\bar 2}}
dz^2 d{\bar z}^2 + 
{\Big (}{dz^3} + a^i dz^i{\Big )}{\Big (}{d{\bar z}^3} +
b^j d{\bar z}^j{\Big )}, }
where $a_i,b_j$ are complex variables that can be obtained from
\sixmetric. The form of the
metric reminds us of the four-dimensional
Taub-NUT space.
The $(3,0)$-form can be constructed
from the negative chirality spinors that are $\cal H$-covariantly
constant
\eqn\threeform{\omega = {1\over 2}~ e^{\alpha \phi}~ \eta^{\top}_{-} \Gamma_{123}
\eta_{-}~ dz^1\wedge dz^2\wedge dz^3, }
where $\alpha$ is a constant. We have to determine this
constant for our case since our conventions
are different to those of \rstrom, and therefore
$\alpha$ will be different. We have an additional
factor of $1/ 2$ in \threeform\ because
for our space $dz^3$
shifted according to
\eqn\sifofzth{
dz^3 ~~\to~~ {1 \over 2}dz^3 + D {\bar z}^2 dz^1 -
A {\bar z}^1 dz^2.}

This $(3,0)$-form
should be independent of background fields. In particular we expect it to
be independent of dilaton, i.e
\eqn\dilindep{ \partial_{\phi} \omega = 0. }
{}From the form of \threeform\ it is not {\it a-priori}
clear how this could be
because the dilaton appears explicitly in the
definition of $\omega$. On the other hand, it is
clear from the set of transformations \rescale\
that the the dilaton dependence enters secretly
after the rescaling
\eqn\gamma{ \Gamma_a~ \to ~e^{\phi\over 4}~\Gamma_a,
\qquad \qquad {\rm and} \qquad
\eta_- ~ \to ~ e^{-{\phi \over 8}} \eta_-.}
Using this we conclude $\alpha = -2$.
The $(3,0)$-form is determined completely in terms of the
complex structure alone.
Because of the compactness of the six-manifold any holomorphic 3-form we
construct in this space would necessarily be a constant multiple of
$\omega$.

Thus once we know the holomorphic form and the complex structure of the
six-manifold we can in principle determine the background $\cal H$ field and
the dilaton $\phi$. However,
so far we have not addressed the crucial question of the existence of this
form. It turns out that if ${\cal H} \ne 0$ but\foot{This eliminates the
familiar Calabi-Yau compactifications where ${\cal H} = d{\cal H} =0$.}
$d{\cal H} = 0$ then there exists no holomorphic
form on the six-manifold \PapaDI. For our case $d{\cal H} \ne 0$
and the non zero
value comes from the additional term that we have in \Gflux. As discussed
earlier this additional term comes from the localized $G$-fluxes near the
singularities. On the type I side these localized fluxes will appear as gauge
fields on the D9-branes which, when $S$-dualised will be the heterotic
gauge fields.

\newsec{Discussion}

In this paper we have given an explicit compactification of the
heterotic string on a six-dimensional manifold which is non-K\"ahler
and has vanishing first Chern class. We have seen that the
3-form flux is non-vanishing.
The metric of the
six-dimensional manifold is determined in \sixmetric\ and is shown
to satisfy the torsional constraints that are imposed by the
form of the supersymmetry transformations.
This explicit formulation of the
background should be useful to study other non-K\"ahler manifolds
with torsion that
can arise by removing some of the restrictions that we have
imposed in section 4.2. The metric for the most generic background
is derived in section 3.3. Showing that the general case
also satisfies the
torsional constraints is mathematically intense and we have not
attempted to do it here. We are implicitly assuming that we
can cancel all the anomalies by fluxes without using any
wandering D3-branes. In the presence of D3-branes the
torsional equations will themselves get modified by the
backreactions from heterotic instantons. A more detailed analysis
is therefore required to verify the torsional constraints
for this case.
But it is clear from duality arguments that this is a valid
supergravity background.

There are other interesting questions and open problems
that need to be
addressed to complete the story:

\item{\tria} From \metnowofti\ the topology of these manifolds
is clear. They have a specific fibration structure. But important
questions like the Euler characteristics or the value of
the Hodge numbers
have to be answered. Since the metric and the topology is known, these
details should not be too hard to determine.

\item{\tria} In \KachruHE\ a type IIB background is determined by
minimizing the superpotential \kstsuper. We also expect the
heterotic background to follow from a superpotential.
The torsional equations \compon\ and the
Donaldson-Uhlenbeck-Yau equation for the gauge bundle
should follow from an effective action in four dimensions.
This matter is under investigation
and more details will appear elsewhere \WIP.

\item{\tria} As briefly discussed in the text, the standard embedding
of the spin connection into the gauge connection
will lead to a trivial dilaton and a constant warp factor \PapaDI.
Therefore, having a non-constant warp factor will depend on the details
of the hermitian structure of our six-manifold and the choice of
gauge connection. This aspect can be used for building models that
are phenomenologically attractive.

\item{\tria} Throughout the paper we have mostly
concentrated on the ${\cal M}$-theory compactification on
$K3 \times K3$.
For the compactification on $T^8/{\cal I}_8$ there
will be no D7-branes or O7-planes
in the type IIB lift of the ${\cal M}$-theory
solution. Therefore, if we cancel the
anomalies completely by fluxes\foot{We thank Shamit Kachru for
discussions on this point.} there will only be an abelian gauge
symmetry in the four-dimensional space-time. It is an interesting
question to trace what really happened to the $SO(32)$ or $E_8
\times E_8$ gauge bundles on the heterotic side.

\item{\tria} Another important question concerns the
size limit of our manifolds. We
showed that for large radii these manifolds become
Calabi-Yau and that this was a contradiction because the
constant total fluxes could not be supported by Calabi-Yau
manifolds. As suggested by \rstrom\ the
3-form equation $dH = \alpha' (p_1(R) - p_1(F))$ has a scale
$\alpha'$ which can be used to fix the radius. This is presently
under investigation \WIP.

\bigbreak\bigskip\bigskip

\centerline{{\bf Acknowledgements}}

\nobreak It is our pleasure to thank E. Bergshoeff, S. Hellerman, C. Hull,
 T. Ortin, J. Polchinski, S. Prokushkin, G. Rajesh, M. Schulz, S. Sethi,
M. M. Sheikh-Jabbari, A. Strominger, E. Witten
and especially M. Becker and S. Kachru for
helpful conversations. The work of K.B. was supported in part
by the University of Utah and the work of K.D. was supported
in part by a David and Lucile Packard
Foundation Fellowship 2000 $-$ 13856.

\vfill

\break

\vfill

\eject

\listrefs

\bye